\documentclass[aps,prx,twocolumn,nofootinbib,10pt]{revtex4-2}
\usepackage{libertine}
\usepackage[T1]{fontenc}
\usepackage[libertine,upint]{newtxmath} 
\usepackage[cal=stixtwofancy,frak=stixtwo]{mathalfa}
\usepackage{graphicx,microtype,mathtools,siunitx,bm,booktabs,xcolor}
\definecolor{linkscolor}{cmyk}{0.6, 0.3, 0, 0.9}
\usepackage[colorlinks=true,allcolors=linkscolor!60]{hyperref}

\DeclareSIUnit\angstrom{\text{Å}}

\DeclarePairedDelimiter\ket{\lvert}{\rangle}
\DeclarePairedDelimiter\av{\langle}{\rangle}
\DeclarePairedDelimiterX\braket[2]{\langle}{\rangle}{#1 \delimsize\vert #2}
\DeclarePairedDelimiterX\matrixel[3]{\langle}{\rangle}{#1 \delimsize\vert #2 \delimsize\vert #3}
\DeclarePairedDelimiter\abs{\lvert}{\rvert}

\DeclareMathOperator{\Tr}{Tr}

\DeclareMathOperator{\eup}{E\!\uparrow}
\DeclareMathOperator{\edn}{E\!\downarrow}
\DeclareMathOperator{\hup}{H\!\uparrow}
\DeclareMathOperator{\hdn}{H\!\downarrow}

\newcommand*{\isotope}[2]{\textsuperscript{#2}#1}
\NewDocumentCommand{\spinor}{m}{%
    \begin{pmatrix} #1 \end{pmatrix}%
}
\NewDocumentCommand{\normord}{m}{%
    :\mathrel{\mspace{1mu} #1 \mspace{1mu}}:%
}
\NewDocumentCommand{\grad}{e{_^}}{%
    \mathop{}\!
    \nabla
    \IfValueT{#1}{_{\!#1}}
    \IfValueT{#2}{^{#2}}
}

\makeatletter
\g@addto@macro\@floatboxreset\centering
\makeatother

\begin{document}

\title{Electron Interactions in Rashba Materials}

\author{Yasha Gindikin}
\author{Alex Kamenev}
\affiliation{W.I.\ Fine Theoretical Physics Institute and School of Physics and Astronomy, University of Minnesota, Minneapolis, Minnesota 55455, USA}

\begin{abstract}
    We present a bunch of novel phenomena stemming from the pair spin-orbit interaction (PSOI), which does not rely on structure inversion asymmetry but instead arises from Coulomb fields of interacting electrons in materials with a strong Rashba effect. First, PSOI can induce $p-$wave superconducting order without the need for any mediators of attraction. Depending on the sign and strength of the PSOI coupling, two distinct superconducting phases emerge in 3D systems, analogous to the A and B phases observed in superfluid \isotope{He}{3}. In contrast, 2D systems exhibit $p^x\pm i p^y$ order parameter, leading to the time-reversal-invariant topological superconductivity. Second, a sufficiently strong PSOI can induce ferromagnetic ordering. It is associated with a deformation of the Fermi surface, which eventually leads to a Lifshitz transition from a spherical to a toroidal Fermi surface, with a number of experimentally observable signatures. Finally, in sufficiently clean Rashba materials, ferromagnetism and $p-$wave superconductivity may coexist. This state resembles the A\textsubscript{1} phase of \isotope{He}{3}, yet it may avoid nodal points due to the toroidal shape of the Fermi surface.
\end{abstract}
 
\maketitle  
\section{Introduction}
The study of spin-orbit interaction remains a fascinating area of research in condensed matter physics for several decades~\cite{10.1063/5.0212170,Bihlmayer2022,manchon2015new}. Rashba spin-orbit interaction (RSOI) offers a new way to control spin states using purely electrical means rather than magnetic fields. This presents a massive advantage for spintronics, a field of study that aims to exploit the quantum spin degree of freedom~\cite{Dyakonov2017,glazov2018electron} to develop new computational and data storage technologies~\cite{Dieny2020,He2022}.

In recent years, the discovery of materials exhibiting giant RSOI has opened up new avenues for investigating exotic quantum phenomena and exploring potential applications in spintronic devices~\cite{PhysRevLett.107.096802,ishizaka2011giant,PhysRevB.95.165444,manzeli20172d,varignon2018new,otrokov2018evidence,PhysRevB.99.085411,PhysRevLett.117.126401}. 
These include the engineered $\mathrm{LaAlO_3/SrTiO_3}$ heterostructures~\cite{doi:10.1021/acs.nanolett.8b01614,Mikheev2023Jun,beyl2015superconducting} and graphene on the two-dimensional (2D) layers of the transition metal dichalcogenides (TMDs)~\cite{Wang2015,Sun_2023,zollner2023twist}, to name a few.

The Rashba effect~\cite{bychkov1984properties}, characterized by locking of spin and momenta directions, results in lifting the spin-degeneracy of conduction band electrons at finite momentum, away from the center of a Brillouin zone. While RSOI may occur from the structure inversion asymmetry, in the present paper we focus exclusively on the \textit{extrinsic} RSOI, produced by electric fields external to the crystalline lattice. Extrinsic RSOI arises due to the non-commutativity of the extrinsic electric potential with the intrinsic Hamiltonian of the host crystal~\cite{Bert}. On a more formal level, it originates from a momentum-dependent Berry curvature, emerging due to a hybridization between the conduction and valence bands. If the valence bands are split by intra-atomic relativistic SOI, the Berry curvature acquires a spin-dependent component, which gives rise to RSOI\@.

The origin of the Berry curvature lies in the non-commutativity of a \textit{momentum-dependent} unitary transformation, $U({\bm{p}})$, from atomic orbitals to Bloch bands basis and a \textit{coordinate-dependent} scalar potential, $\varphi({\bm{r}})$. It modifies dynamics of quasiparticles, restricted to a occupy states in a specific band, under the influence of an external electric field. The Moyal expansion of these non-commutative operators leads to gradient corrections of the form $\varepsilon_{\lambda\mu\nu}\grad^\nu_{\bm{r}}\varphi({\bm{r}})\grad^\mu_{\bm{p}}U(\bm{p})\,\hat\sigma^\lambda$, where $\mu,\nu,\lambda$ are Cartesian coordinates and $\hat \sigma$ are spin Pauli matrices.  In this way Rashba spin-orbit effects are inextricably proportional to an extrinsic electric field  $\bm{\mathcal{E}}(\bm{r})= - \bm{\nabla}_{\! \bm{r}}\varphi({\bm{r}})$.   

For conduction band electrons near the Brillouin zone center, the resulting coupling can be effectively described by the widely-used \textit{single-particle} Rashba Hamiltonian, which is linear in the wave vector, $\bm{p}$,~\cite{winkler}
\begin{equation} 
    \label{Eq:H-Rashba}
    H_{\mathrm{RSOI}} = \alpha \left[\bm{\mathcal{E}}\times {\bm{p}}\right]\cdot \hat{\bm{\sigma}},
\end{equation} 
where $\alpha$ is a material-dependent Rashba constant. In 2D systems, the extrinsic electric field, $\bm{\mathcal{E}}$, can originate from external factors such as gate voltages. In contrast, in 3D metals, uniform electric fields are typically screened. However, local fields can arise from charged impurities and structural defects, giving rise to extrinsic RSOI, which leads to skew scattering and side-jumping of electrons~\cite{SMIT195839,Perel,PhysRevB.2.4559,Nozieres}. These mechanisms have since been experimentally validated and are now recognized as significant contributors to phenomena such as the anomalous Hall and spin Hall effects. Notably, these effects have been observed in a wide range of materials, including those of current interest, such as topological insulators and 2D materials~\cite{RevModPhys.82.1539,RevModPhys.87.1213}.

The notion of pair spin-orbit interactions (PSOI)~\cite{2019arXiv190506340G} represents a conceptual shift from these traditional SOI mechanisms. Unlike the single-particle RSOI, which arises from external electric fields, PSOI originates directly from the Coulomb fields of conduction band electrons. Therefore, PSOI does \textit{not} rely on structural inversion asymmetry. This key insight — that the electric field, $\bm{\mathcal{E}}$, in Eq.~\eqref{Eq:H-Rashba} can arise from the Coulomb forces — opens up an entirely new research avenue. This coupling directly affects the two-body interactions between electrons, rendering them both spin- and momentum-dependent. As a result, the modified two-body interactions lead to phenomena, distinctly different from both the conventional single-body RSOI and usual two-body interactions.

It was established that PSOI can significantly impact the charge susceptibility, giving rise to instabilities in the \textit{particle-hole} channel in low-dimensional electron systems. In 1D systems, PSOI breaks the spin-charge separation, a hallmark of Luttinger liquid theory, leading to a correlated state with distinctive signatures in electron transport~\cite{PhysRevB.95.045138,2017arXiv170700316G,doi:10.1002/pssr.201700256}. When PSOI exceeds a critical value, it triggers quantum phase transitions in both 1D and 2D systems, where a non-trivial charge order can emerge~\cite{GINDIKIN2022115328,PhysRevB.109.075163}.

On the other hand, the nature and consequences of PSOI-induced instabilities in the \textit{particle-particle} channel remain largely unexplored. To date, only solutions to the two-body Schrödinger equation have been obtained, demonstrating that sufficiently strong PSOI can lead to the formation of bound electron pairs with a particular orbital and spin structures, including $p$-wave symmetry~\cite{PhysRevB.98.115137,10.1016/j.physe.2018.12.028}, and may even support finite-momentum pairing~\cite{2018arXiv180410826G,PhysRevB.98.115137} and bound states in continuum~\cite{2019arXiv190409510G}.~\footnote{The formation of bound pairs due to PSOI is just one aspect of a broader phenomenon in many-band systems, where bound electron pairs with energies within the band gap can emerge. Such pairs have been studied in graphene, bilayer graphene~\cite{Sabio2010,MarnhamShytovPRB2015}, flat-band systems~\cite{Torma,Iskin}, Dirac semimetals~\cite{Portnoi}, topological insulators~\cite{PhysRevB.95.085417}, etc~\cite{perel1971}. Two primary mechanisms for effective electron attraction in these systems have been identified: the formation of a negative effective mass due to subband mixing~\cite{PhysRevB.95.085417,keldysh1964deep}, and attractive PSOI induced by Coulomb interactions in Rashba materials~\cite{bhzbeps}. Both mechanisms represent different facets of subband hybridization.}
Exploring bound pairs and larger few-body complexes~\cite{Kezerashvili2019,https://doi.org/10.1002/qua.25994,10.1093/oxfmat/itad004} is a logical first step toward solving the much more intricate problem of correlated many-electron states. Indeed, the \textit{collective behavior} of electron systems with PSOI remains an open question. Specifically, how PSOI-induced instabilities in the particle-particle channel stabilize, and whether they lead to superconductivity or alternative phases, remains unanswered. The identification of relevant order parameters and solutions of the corresponding self-consistency equations are required to obtain the ground state of the many-electron system with PSOI\@. 

The present work aims at making first steps in this direction. To this end, it investigates the superconducting and magnetic instabilities in both 3D and 2D Fermi liquids with PSOI\@. We demonstrate that PSOI can induce a $p-$wave superconducting state in 3D, which can be either nodal or nodeless depending on the sign of the SOI constant, $\alpha$. The order parameters of these superconducting phases are similar to the A and B phases of \isotope{He}{3}~\cite{wolfle}. Notably, these superconducting states do not require phonons or any other mediators of attraction, relying solely on the interplay between Coulomb repulsion and PSOI\@. In 2D systems, the $p-$wave superconducting states are fully gapped, pairing spin-up and spin-down electrons into $p^x\pm i p^y$ states, correspondingly. The resulting superconducting state respects both time-reversal and particle-hole symmetries, putting it into DIII symmetry class~\cite{haim2019time}, topological in 2D\@.

While electron pairing has also been predicted within 2D Rashba-Hubbard models, the mechanism there is based on Kohn-Luttinger effect, where Rashba SOI mainly induces a mixing of spin-singlet and spin-triplet channels~\cite{PhysRevB.93.214516,PhysRevLett.120.177002,PhysRevB.102.174512,doi:10.1143/JPSJ.77.124711,doi:10.7566/JPSJ.82.014702}. In contrast, our work introduces a conceptually different mechanism of superconducting pairing. This mechanism is a direct, first-order effect in the interaction strength, unlike the much weaker, second-order Kohn-Luttinger mechanism.

The PSOI also facilitates and affects a ferromagnetic (FM) order coming from the Bloch-Stoner  mechanism. Due to PSOI the FM breaking of the rotation symmetry in the spin space is associated  with a deformation of the Fermi surface in the momentum space. Eventually such a deformation leads to a topology-changing Lifshitz transition from a genus-zero (simply connected) to a genus-one (toroidal-shaped) Fermi surface. This transition has distinct observable signatures in material's thermodynamic and transport properties. Specifically, it leads to a changes in Shubnikov-de Haas and Friedel oscillations patterns, which can serve as direct probes of the Fermi surface topology. Moreover, a potential coexistence of FM and $p-$wave superconducting orders, which has garnered considerable attention in the field~\cite{HUXLEY2015368,liu2021iron,saxena2000superconductivity,ran2019nearly,hu2024possible}, presents an intriguing scenario. Interestingly, the toroidal Fermi surface allows for a node-less fully spin-polarized order parameter.  

\section{Electron-electron interactions in Rashba Materials}

In this section we work out consequences of the phenomenological Rashba spin-orbit coupling, Eq.~\eqref{Eq:H-Rashba}, for electron-electron interaction Hamiltonian. To this end consider a 3D Rashba system without structure inversion asymmetry. A 2D case is a simple particular case of this consideration, which is briefly mentioned in the end. We do not consider the effects of disorder in our analysis. Furthermore, since a bulk electric field is not allowed in a 3D electronic system, it might be tempting to conclude that the Rashba spin-orbit interaction term, as given in Eq.~\eqref{Eq:H-Rashba}, is inconsequential. However, such conclusion is premature because of the fluctuating electric fields created by electron-electron interactions. For weakly screened Coulomb interactions, considered below, these fluctuating Coulomb fields can reach magnitudes on the order of $10^7$--$10^8$ V/cm, which are comparable to the built-in electric fields responsible for giant Rashba splitting in van der Waals materials~\cite{manchon2015new}.

\subsection{Single-band picture of electron dynamics}

To describe the physics originating from this observation one deals with a two-component spinor electron creation operator $\hat{\psi}_{\bm{p}}^{\dagger} \equiv \spinor{\hat{c}_{\bm{p},\uparrow}^{\dagger} \hat{c}_{\bm{p},\downarrow}^{\dagger}}$, describing conduction band electrons in a spin-up/spin-down basis.  The corresponding electron creation operator in the coordinate basis is denoted as $\hat{\psi}^{\dagger}(\bm{r})$.

It is convenient to define the spin current \textit{tensor} as
\begin{equation}
    \label{Eq:SpinCurrentTensor}
    \hat{\jmath}^{\lambda \mu}(\bm{r}) = \frac{1}{2m i} \left( \hat{\psi}^{\dagger}(\bm{r}) \hat{\sigma}^\lambda \grad_{\bm{r}}^{\mu}{\hat{\psi}(\bm{r})}  - (\grad_{\bm{r}}^{\mu}{\hat{\psi}^{\dagger}(\bm{r})})\hat{\sigma}^\lambda \hat{\psi}(\bm{r}) \right),
\end{equation}
with Greek indices $\lambda, \mu, \ldots$ standing for the Cartesian components.\footnote{Throughout the text we set $\hbar = 1$, and suppress the ${(volume)}^{-1}$ factors before the sums over momenta.} This object describes a flow of the $\lambda$ component of the spin angular momentum in the $\mu$ spatial direction. 
The Hamiltonian for electrons interacting with an external scalar electric potential, $\varphi(\bm{r},t)$, which includes RSOI, Eq.~\eqref{Eq:H-Rashba}, is  
\begin{equation}
            \label{Eq:H-int-coordinate}
    \begin{split}
        \hat{H}_{\mathrm{RSOI}} &= -   \int d\bm{r}\,\left(e\varphi(\bm{r},t) \hat{\rho}(\bm{r}) + \alpha m \,\varepsilon_{\lambda\mu\nu} \mathcal{E}^\nu(\bm{r}) \hat{\jmath}^{\lambda\mu}(\bm{r}) \right)  \\
        &= -   \int d\bm{r}\, \varphi(\bm{r},t)\left( e\hat{\rho}(\bm{r}) + \alpha m\,  \varepsilon_{\lambda\mu\nu} \grad_{\bm{r}}^{\nu} \hat{\jmath}^{\lambda\mu}(\bm{r}) \right)\\
        &= -   \int d\bm{r}\, \varphi(\bm{r},t)\left( e \hat{\psi}^{\dagger} \hat{\psi} + i\alpha\,  \varepsilon_{\lambda\mu\nu} \grad_{\bm{r}}^{\mu} \hat{\psi}^{\dagger} \hat{\sigma}^\lambda
        \grad_{\bm{r}}^{\nu} \hat{\psi}  \right),
    \end{split}
\end{equation} 
where $\hat{\rho}(\bm{r}) \equiv \hat{\psi}^{\dagger}(\bm{r})\hat{\psi}(\bm{r})$ is the electron density, and $\varepsilon_{\lambda\mu\nu}$ is the Levi-Civita symbol; summation over the repeated indices is implied. In the second line we used that $\mathcal{E}^\nu(\bm{r}) = -\grad_{\bm{r}}^{\nu} \varphi(\bm{r})$ and performed integration by parts. 
In the same order in the gradient expansion there is one extra symmetry-allowed contribution, known in relativistic quantum mechanics as the Darwin term~\cite{akhiezer1957quantum},
\begin{equation}
    \label{Eq:Darwin}
    \hat H_{D}= -\frac{\beta}{2}  \int d\bm{r} \grad_{\bm{r}}^{\mu}\grad_{\bm{r}}^{\mu} \varphi(\bm{r},t) \, \hat{\psi}^{\dagger}  \hat{\sigma}_0\, \hat{\psi}\,.
\end{equation}

Appearance of both Rashba and Darwin terms warrants some elaboration. The very notion of the SOI is an attribute of a simplified single-band portrayal of the electron dynamics, accomplished via the projection of a microscopic multi-band Hamiltonian onto the conduction band. Through this projection, an effective Hamiltonian emerges with the SOI and Darwin terms, reflecting the contribution of the valence band(s) to the conduction band electrons dynamics. In particular, the local scalar potential disturbs both valence and conduction bands.  Due to the momentum-dependent nature of the rotation from local orbitals to the band representation, it leads to the gradient correction to the conduction-band density response. The RSOI term in Eq.~\eqref{Eq:H-int-coordinate} reflects the spin-dependent part of this gradient correction, while the Darwin term~\eqref{Eq:Darwin} — the spin-independent one. 

Within the  Kane $8 \times 8$ model of the semiconductor band structure~\cite{voon2009kp}, the corresponding constants are given by~\cite{PhysRevB.47.16020,winkler}
\begin{equation}
\label{Eq:Kane_constants}
    \alpha = \!\frac{e P^2}{3}\!\left[ \frac{1}{{(E_0+ \Delta_{SO})}^2} - \frac{1}{E_0^2}\right];\,\, 
  \beta = \!\frac{e P^2}{3}\!\left[ \frac{1}{{(E_0+ \Delta_{SO})}^2} + \frac{2}{E_0^2}\right],  
\end{equation}
with the energy gap $E_0$, the energy of the split-off holes $\Delta_{SO}$, and the dipole matrix element $P$. One may notice that the constants satisfy the following inequalities,\footnote{The sign of the Darwin correction as given in book~\cite{winkler} should be reversed.}
\begin{equation}
    \label{Eq:inequalities}
   \beta>0;\quad\quad -\beta/2 \le \alpha \le \beta. 
\end{equation}
These bounds are actually more general than the specific model, and may be shown to be valid for any rotationally invariant multi-band model with linear-in-momentum hybridization. The lower bound, $\alpha=-\beta/2$, is reached in the Kane model with an infinitely large spin-splitting of the valence bands, while the upper bound, $\alpha=\beta$, corresponds to the Dirac model. 

We emphasize, however, that materials with a giant Rashba effect, such as graphene on TMDs, or Van der Waals materials with heavy adatoms, are not known to be described by Kane-like models. We thus consider the corresponding coupling constants as phenomenological parameters, which are not necessarily bound by the inequalities~\eqref{Eq:inequalities}.

One may be surprised that there is no symmetry between positive and negative $\alpha$ and that the sign of the SOI strength, $\alpha$, in Eq.~\eqref{Eq:H-int-coordinate} does have a physical significance. Indeed, one may think that  the change of $\alpha\to -\alpha$ may be always compensated by $\bm{\sigma}\to -\bm{\sigma}$ transformation. Yet this logic is erroneous, since   $\bm{\sigma}\to -\bm{\sigma}$ is not a faithful representation of $\uparrow\leftrightarrow \downarrow$ exchange. Indeed, consider $\hat{\psi}^{\dagger} \hat{\sigma}^\lambda \hat{\psi} $ combination. While for $\lambda=y,z$ one indeed observes that $\uparrow\leftrightarrow \downarrow$ is equivalent to $\sigma^{y,z}\to -\sigma^{y,z}$, this logic does not work for $\lambda=x$. Therefore, if all three Pauli matrices are employed (as in 3D case), $\bm{\sigma}\to -\bm{\sigma}$ is not a legitimate transformation and a sign of $\alpha$ does have physical significance, affecting e.g.\ the spin texture of the electron state~\cite{PhysRevB.84.115426,PhysRevB.91.035403}. On the other hand, if SOI term is limited to two Pauli's (like in 2D, where $\lambda =z$ in Eq.~\eqref{Eq:H-int-coordinate}, or $\lambda=x,y$ in Eq.~\eqref{Eq:H-Rashba}), the physical observables are invariant with respect to $\alpha\to -\alpha$. The actual sign of the Rashba SOI in realistic structures depends on a particular charge asymmetry near the atomic cores~\cite{MANCHON20171}.

In the momentum representation Eqs.~\eqref{Eq:H-int-coordinate} and~\eqref{Eq:Darwin} result in the following interaction vertex between the conduction band electrons and the external scalar potential 
\begin{equation}
    \hat{H}_{\mathrm{ext}} = \sum_{\bm{p}_1,\bm{p}_2}
    \varphi(\bm{p}_2-\bm{p}_1,t) \, 
    \hat{\psi}^{\dagger}_{\bm{p}_1} \hat{\Gamma}_{\bm{p}_1\bm{p}_2} \hat{\psi}_{\bm{p}_2}\,, 
\end{equation} 
where the vertex $\hat{\Gamma}_{\bm{p}_1\bm{p}_2}$ is a matrix in the spin as well as in the momentum spaces given by
\begin{equation}
    \label{Eq:VertexMomentum}
    \hat{\Gamma}_{\bm{p}_1\bm{p}_2} = -e \hat{\sigma}_0  +\frac{\beta}{2}{(\bm{p}_1 -\bm{p}_2)}^2\hat{\sigma}_0 - i \alpha [\bm{p}_1 \times \bm{p}_2]\cdot \hat{\bm{\sigma}} \,. 
\end{equation}
Notice that the anomalous contributions to the vertex affect only the local distribution of the electron density, while the total number of electrons in the conduction band is a constant. This is evident from the fact that both Rashba and Darwin terms vanish in the long-wave limit where $\bm{q} = \bm{p}_1 - \bm{p}_2 \to 0$.

\subsection{Pair spin-orbit interaction}

In interacting systems the scalar potential is a quantized Gaussian fluctuating field with the correlation function given by  $\av{ \varphi(\bm{q},t)\varphi(-\bm{q},t')} \equiv \mathcal{U}_{\bm{q}}\, \delta(t-t')$. In 3D materials the latter is the usual Coulomb potential, $\mathcal{U}_{\bm{q}}=4\pi/q^2$, while in 2D it may include effects of dielectric constant $\epsilon$ of a surrounding media and, possibly, screening by external metallic gates.  This leads to the e-e interaction Hamiltonian of the form:
\begin{equation}
    \label{Eq:HamInt}
    \hat{H}_{\mathrm{int}} = \frac{1}{2} \sum_{\substack{\bm{p}_1,\bm{p}_2\\\bm{p}'_1,\bm{p}'_2}}
    \mathcal{U}_{\bm{p}_2-\bm{p}_1}\!  
    \normord{\hat{\psi}^{\dagger}_{\bm{p}_1} \hat{\Gamma}_{\bm{p}_1\bm{p}_2} \hat{\psi}_{\bm{p}_2} \hat{\psi}^{\dagger}_{\bm{p}'_1} \hat{\Gamma}_{\bm{p}'_1\bm{p}'_2} \hat{\psi}_{\bm{p}'_2}}\,,
\end{equation} 
where the momentum summation is limited to momentum conserving processes with $\bm{p}_1+\bm{p}'_1=\bm{p}_2+\bm{p}'_2$ and the colon signs denote the normal ordering of the electron creation/annihilation operators.~\footnote{While there is a formal similarity between the vertex structure here and that in the context of fluctuating Rashba SOI mediated by soft phonons near the ferroelectric transition in $\mathrm{SrTiO_3}$~\cite{PhysRevB.101.174501}, we emphasize that PSOI is mediated by screened Coulomb interactions and does not involve the crystalline lattice. This key difference makes our mechanism potentially applicable to a wider range of materials, where phonon-mediated interactions might coexist but are not a prerequisite.}

The total Hamiltonian $ \hat{H}_{\mathrm{tot}} =  \hat{H}_{\mathrm{kin}} +  \hat{H}_{\mathrm{int}}$ also includes the kinetic energy
\begin{equation}
    \hat{H}_{\mathrm{kin}} = \sum_{\bm{p}}  \hat{\psi}^{\dagger}_{\bm{p}} \, \xi_{\bm{p}}\hat{\sigma}_0\, \hat{\psi}_{\bm{p}}\,.
\end{equation}
To be specific, we assume quadratic energy dispersion, $\xi_{\bm{p}} = \bm{p}^2/2m-\mu$. 

Let us emphasize  a distinctive aspect of the electron-electron interactions in Rashba materials. In the conventional scenario the \textit{e-e} interaction is determined by the Coulomb contribution, $e^2\mathcal{U}_{\bm{q}}$, only. In this case the interaction strength, defined by the ratio of the interaction energy to the Fermi energy, is characterized by a single dimensionless parameter, $ r_s $, — the mean inter-electron distance normalized to the Bohr radius $ a_B $; one has $r_s={(\gamma_d k_F a_B)}^{-1}$, with $\gamma_3=\sqrt[3]{4/(9\pi)}$ for 3D systems, and $\gamma_2=1/\sqrt{2}$ for 2D~\cite{Vignalebook}. The PSOI is characterized by another interaction parameter, $\tilde{\alpha}/r_s$, where $\tilde{\alpha}$ is a dimensionless SOI constant expressed in atomic units as $\tilde{\alpha} = \alpha/(e a_B^2)$~\cite{review_jetp}. Similarly, the Darwin interaction parameter is given by $\tilde{\beta}/r_s$, with $\tilde{\beta} = \beta/(e a_B^2)$. We summarize these observations in the table:
\begin{center}
    \begin{tabular}{cccc}
        Interaction: & \quad Coulomb &  \quad PSOI & \quad Darwin\\
        \midrule
          $ \dfrac{E_{\mathrm{int}}}{E_{\mathrm{kin}}} $ & \,$ r_s $ & \,\,\,$ \dfrac{\tilde{\alpha}}{r_s} $ &\,\,\, $ \dfrac{\tilde{\beta}}{r_s} $
    \end{tabular}
\end{center}
The normal and anomalous interactions thus exhibit qualitatively different dependencies on the electronic system's characteristics. Notably, while the Coulomb interaction parameter $ r_s $ decreases with increasing electron density, the PSOI parameter $ \tilde{\alpha}/r_s $ increases as the electron density increases.

This implies that PSOI effects become increasingly significant in dense electron gases, where $ r_s < 1 $. This is the regime, where ordered and unconventional superconducting phases may be expected to show up in Rashba systems~\cite{GINDIKIN2022115328,PhysRevB.109.075163}. At the same time, the effects of Thomas-Fermi (TF) screening are minimal at low $r_s$, reflected in the small ratio of the TF inverse screening radius  $ \kappa $ to the Fermi momentum $ k_F $, with $ \kappa^2/k_F^2 \propto r_s $. Consequently, the long-range nature of the e-e interaction potential must be explicitly taken into account.

However, the PSOI parameter cannot increase indefinitely with the electron density. A very dense electron gas behaves nearly as a non-interacting. At extremely high densities, when the Fermi energy approaches the energy gap to other Bloch bands contributing to the SOI, the assumptions of the single-band model of SOI, built upon the $ k\cdot p $ approximation, are invalidated. This places a practical lower limit on the value of $ r_s $, dictated by the specific material's band structure.

To accommodate these considerations, the theory of PSOI must be generalized to include momentum dependence in the SOI ``constant,'' $ \alpha \to \alpha_{p_1 p_2} $, allowing $ \alpha_{p_1 p_2} $ to decrease when either momentum exceeds a certain threshold. A particular form of such dependence can be obtained from a more detailed microscopic analysis within a multi-band model, as outlined in Appendix~\ref{Sec:Microscopics}, or from density functional theory (DFT) calculations~\cite{PhysRevResearch.5.023177}.

In 2D systems, one may also take into account the one-body RSOI produced by the uniform normal electric field $\mathcal{E}_z$ of the gate. In this case, the single-particle Hamiltonian is
\begin{equation}
    \hat{H}_{\mathrm{kin}} = \sum_{\bm{p}}  \hat{\psi}^{\dagger}_{\bm{p}} 
    \left( \xi_{\bm{p}} \hat{\sigma}_0+ \alpha\, \mathcal{E}_z {[\bm{p}\times\hat{\bm{\sigma}}]}_z\right)
    \hat{\psi}_{\bm{p}}\,,
\end{equation}
where $\bm{p}$ is a 2D vector in $x-y$ plane. The corresponding interaction vertex acquires the spin-diagonal form:
$\hat{\Gamma}_{\bm{p}_1\bm{p}_2} = -e \hat{\sigma}_0  +\tfrac{\beta}{2}(\bm{p}_1 -\bm{p}_2)^2\hat{\sigma}_0 - i \alpha{[\bm{p}_1 \times \bm{p}_2]}_z \hat{{\sigma}}_z$, as both the initial and final momenta are in the $x-y$ plane.  

\section{\textit{P-}wave Superconductivity}
\label{sec:p-wave}

Here, we show that the interplay of PSOI and Coulomb repulsion favors $p-$wave superconducting order, circumventing the need for phonons or other mediators of attraction. The mechanism bears similarities to the Kohn-Luttinger higher angular momentum superconductivity~\cite{PhysRevLett.15.524,Chubukov2022}. The key distinction lies in the source of the effective attraction in high angular channels. In the present case it originates from the PSOI\@. Notably, our mechanism is a direct, first-order effect in the interaction strength, distinguishing it from the Kohn-Luttinger scenario that arises as a higher-order effect --- second order in 3D and third order in 2D~\cite{PhysRevB.48.1097}.

\subsection{Variational considerations}

Consider the effective Bardeen–Cooper–Schrieffer (BCS) Hamiltonian that describes the pairing interaction between electrons with zero total momentum. Making the spin indexes explicit, the interaction Hamiltonian~\eqref{Eq:HamInt} acquires the form 
\begin{equation}
    \label{Eq:HamBCS}
    \hat{H}_{\mathrm{BCS}} \propto
    \mathcal{U}_{\bm{p}-\bm{p}'} 
\hat{\Gamma}_{\bm{p}\bm{p}'}^{ss'} \hat{\Gamma}_{-\bm{p}-\bm{p}'}^{rr'}\,
{\psi}^{\dagger}_{\bm{p} s} {\psi}^{\dagger}_{-\bm{p} r} {\psi}_{-\bm{p}' r'}  {\psi}_{\bm{p}' s'}, 
\end{equation} 
where we omitted momentum and spin summation symbols for brevity.

Anticipating the spin-triplet superconductivity,\footnote{One can check that the interaction in the $s-$wave channel remains repulsive.} we introduce the anomalous expectation value as a symmetric complex matrix in the spin space:
\begin{equation}
    \label{Eq:anomalous-1}
 \hat{d}^{\,rs}_{\bm{p}}\! =   \av*{{\psi}_{-\bm{p} r}  {\psi}_{\bm{p} s}} = \! \begin{pmatrix}
                                           -d^x_{\bm{p}}+id^y_{\bm{p}} & d^z_{\bm{p}}\\ 
                                           d^z_{\bm{p}} & \!\!\!\!d^x_{\bm{p}}+id^y_{\bm{p}}
                                           \end{pmatrix}\! 
=\bm{d}_{\bm{p}}\cdot {(\bm{\hat{\sigma}}\, i\hat{\sigma}_y)}_{rs}\,.  
\end{equation} 
The combinations $\mp d^x_{\bm{p}}+id^y_{\bm{p}}$ describe $m=\pm 1$ components of the triplet order parameter, while $d^z_{\bm{p}}$ is its $m=0$ component. Here $m$ labels projections of the $l=1$ spin angular momentum.

The expectation value of the BCS Hamiltonian~\eqref{Eq:HamBCS} equals
\begin{equation}
\label{Eq:HamBCS-expectation}
    \av{\hat{H}_{\mathrm{BCS}}}\! \propto\!
    \mathcal{U}_{\bm{p}-\bm{p}'} \!\! 
    \Tr\! \left\{\! (-i\hat{\sigma}_y \bm{\hat{\sigma}})\!\cdot\! \bar{\bm{d}}_{\bm{p}}  \hat{\Gamma}_{\bm{p}\bm{p}'} 
    \bm{d}_{\bm{p}'}\!\cdot\! (\bm{\hat{\sigma}}\, i\hat{\sigma}_y)
    \hat{\Gamma}_{\bm{p}\bm{p}'}^{\intercal} \! \right\},  
\end{equation} 
where the trace and matrix transposition are performed in the spin space only, and we used that $\hat{\Gamma}_{-\bm{p}-\bm{p}'} = \hat{\Gamma}_{\bm{p}\bm{p}'}$. Employing that $\hat\sigma_y{(\hat\sigma^\lambda)}^\intercal \hat\sigma_y = -\hat\sigma^\lambda$ and 
\begin{align*}
    &\Tr \left\{ {\hat{\sigma}}^\mu   {\hat{\sigma}}^\nu \right\}=2\delta^{\mu\nu}; \qquad
    \Tr \left\{ {\hat{\sigma}}^\mu  \hat{\sigma}^\lambda  
    {\hat{\sigma}}^\nu \right\}=2i\varepsilon^{\mu\lambda\nu}; 
    \\
    &\Tr \left\{ {\hat{\sigma}}^\mu  \hat{\sigma}^\lambda  
    {\hat{\sigma}}^\nu \hat{\sigma}^\varsigma \right\} =2\left(\delta^{\mu\lambda}\delta^{\nu\varsigma} -\delta^{\mu\nu}\delta^{\lambda\varsigma} + \delta^{\mu\varsigma}\delta^{\lambda\nu}\right)\,,
\end{align*}
one obtains
\begin{widetext}
    \begin{equation}
    \label{Eq:HamBCS-expectation-11}
        \av*{\hat{H}_{\mathrm{BCS}}} = 2\! \sum_{\bm{p},\bm{p}'} \mathcal{U}_{\bm{p}-\bm{p}'} 
        \big\{ \breve{e}^2 (\bar{\bm{d}}_{\bm{p}}\! \cdot\bm{d}_{\bm{p}'})
        + 2 \breve{e}\alpha\, [\bar{\bm{d}}_{\bm{p}}\! \times \bm{d}_{\bm{p}'}]\cdot \bm{\mathcal{M}}
        + 2\alpha^2 (\bar{\bm{d}}_{\bm{p}}\! \cdot \bm{\mathcal{M}} )( \bm{\mathcal{M}} \cdot {\bm{d}}_{\bm{p}'}) - \alpha^2 (\bar{\bm{d}}_{\bm{p}}\! \cdot\bm{d}_{\bm{p}'}) \mathcal{M}^2 \big\}\,,
    \end{equation} 
\end{widetext}
with $\bm{\mathcal{M}} = [\bm{p} \times \bm{p}']$ and $\breve{e} = e-\tfrac{\beta}{2}(\bm{p}-\bm{p}')^2$.

For the $p-$wave pairing, the odd parity of the vector $\bm{d}_{\bm{p}}$ as a function of the orbital momentum $\bm{p}$ allows for the introduction of a local tensorial order parameter $\hat{D}$ via
\begin{equation} 
    \label{Eq:tensorial_order_parameter}
    d_{\bm{p}}^{\nu} = D_{\nu\nu'} p^{\nu'}.
\end{equation}
Generally speaking, ${D}_{\nu\nu'}$ is a function of the absolute value of momenta, $\abs{\bm{p}}$, which 
may be put to be the Fermi momentum for consideration of low-energy phenomena.  In terms of the tensor ${D}_{\nu\nu'}$, the BCS energy~\eqref{Eq:HamBCS-expectation-11} represents a quadratic form,
\begin{equation}
    \av*{\hat{H}_{\mathrm{BCS}}} = \bar{D}_{\mu \mu'} T^{\mu\mu',\nu\nu'} D_{\nu\nu'}\,,
\end{equation}
labelled by the 9-component index $\nu\nu'=(xx,xy,\ldots,zz)$. The symmetric $9\times 9$ matrix, $T^{\mu\mu',\nu\nu'}$, is obtained via integration over angles in Eq.~\eqref{Eq:HamBCS-expectation-11}.

Due to the rotational symmetry of the e-e interaction potential, the result of the angular integration takes on a particularly simple form. This symmetry ensures that the spherical harmonics of the integrands in Eq.~\eqref{Eq:HamBCS-expectation-11}, representing products of $\mathcal{U}_{\bm{p}-\bm{p}'}$ and rotationally invariant combinations of momenta, are determined by merely a few constants:
\begin{widetext}
    \begin{equation}
    \begin{split}
        \label{Eq:spherical_momenta}
        &\av*{2\mathcal{U}_{\bm{p}-\bm{p}'} \left( \breve{e}^2 - \alpha^2 \mathcal{M}^2 \right) p_{\mu} p'_{\mu'}}_{\theta,\phi} = U \delta_{\mu \mu'}\\
        &\av*{4 \breve{e} \alpha \mathcal{U}_{\bm{p}-\bm{p}'} p_{\mu}p_{\nu} p'_{\mu'}p'_{\nu'}}_{\theta,\phi} = (\tilde{V}+V) \delta_{\mu \nu}\delta_{\mu' \nu'} + (\tilde{V}-V) \left(\delta_{\mu \mu'}\delta_{\nu \nu'}+\delta_{\mu \nu'}\delta_{\nu \mu'}\right)\\
        &\av*{4\alpha^2\mathcal{U}_{\!\bm{p}-\bm{p}'} p_{\mu}p_{\nu}p_{\lambda}p'_{\mu'}p'_{\nu'}p'_{\lambda'}}_{\theta,\phi} = (\tilde{W}+W)\big(\delta_{\mu\mu'}\delta_{\nu\lambda}\delta_{\nu'\lambda'}+\delta_{\nu\mu'}\delta_{\mu\lambda}\delta_{\nu'\lambda'}+\delta_{\lambda\mu'}\delta_{\mu\nu}\delta_{\nu'\lambda'}\\
        &+\delta_{\mu\nu'}\delta_{\nu\lambda}\delta_{\mu'\lambda'}+\delta_{\nu\nu'}\delta_{\mu\lambda}\delta_{\mu'\lambda'}+\delta_{\lambda\nu'}\delta_{\mu\nu}\delta_{\mu'\lambda'}+\delta_{\mu\lambda'}\delta_{\nu\lambda}\delta_{\mu'\nu'}+\delta_{\nu\lambda'}\delta_{\mu\lambda}\delta_{\mu'\nu'}+\delta_{\lambda\lambda'}\delta_{\mu\nu}\delta_{\mu'\nu'}\big)\\
        &+(\tilde{W}-W)\left(\delta_{\mu\mu'}\delta_{\nu\nu'}\delta_{\lambda\lambda'} + \delta_{\mu\mu'}\delta_{\nu\lambda'}\delta_{\lambda\nu'}
        {}+ {} \delta_{\mu\nu'}\delta_{\nu\mu'}\delta_{\lambda\lambda'} + \delta_{\mu\nu'}\delta_{\nu\lambda'}\delta_{\lambda\mu'}
        {}+ {} \delta_{\mu\lambda'}\delta_{\nu\mu'}\delta_{\lambda\nu'} + \delta_{\mu\lambda'}\delta_{\nu\nu'}\delta_{\lambda\mu'}\right)\,.
    \end{split}
    \end{equation}
    Of these, only three contribute to the expression for the matrix $T^{\mu\mu',\nu\nu'}$:
    \begin{equation}
        T^{\mu\mu',\nu\nu'} = (U+4W)\delta_{\mu\nu}\delta_{\mu'\nu'}+2V \varepsilon^{\mu\nu\lambda}\varepsilon^{\mu'\nu'\lambda} +2W \left(\varepsilon^{\mu\mu'\lambda}\varepsilon^{\nu\nu'\lambda} + \varepsilon^{\mu\nu'\lambda}\varepsilon^{\nu\mu'\lambda}\right)\,,
    \end{equation}
\end{widetext}
while $\tilde{V}$ and $\tilde{W}$ cancel out. Its eigensystem includes nine states with eigenvalues given by:
\begin{equation}
    \label{Eq:9x9_eigenvalues}
    \{ \underbrace{U + 4 V}_{\text{singlet}}, \underbrace{U + 2 V + 10 W}_{\text{triplet}}, \underbrace{U - 2 V + 6 W}_{\text{quintet}}\}\,,
\end{equation}
where singlet, triplet, and quintet indicate the 1,3, and 5-fold degeneracy of the corresponding eigenvalues. 
The non-degenerate singlet state is isotropic, with the corresponding eigen-tensor $D_{\nu\nu'}\sim \delta_{\nu\nu'}$, and thus $\bm{d_p}\sim \bm{p}$. The triplet and quintet states break rotational symmetry, since a definite momentum projection along some axis implies the existence of a corresponding rotation plane. The $p-$wave superconducting instability is determined by the most negative of eigenvalues~\eqref{Eq:9x9_eigenvalues}.  

To proceed one needs a specific form of the e-e interaction potential, $\mathcal{U}_{\bm{q}}$. We limit ourselves with the statically screened Coulomb interaction: 
\begin{equation}
\mathcal{U}_{\bm{q}} = \begin{dcases}
                        \frac{4 \pi e^2}{\bm{q}^2+ \kappa^2},\; \quad \kappa^2 = 4\pi e^2 \nu_3; & 3D\\
                        \frac{2 \pi e^2}{\abs{\bm{q}}+ \kappa},\; \quad \kappa = 2\pi e^2 \nu_2; & 2D
                       \end{dcases}
\end{equation}
where $\nu_d$ is the density of states in dimension $d$ at the Fermi energy. 

First, we focus on the case of a weakly screened Coulomb interaction, when PSOI effects are anticipated to be most significant~\cite{review_jetp}. This case corresponds to $\kappa^2/ k_F^2 < 1$, or, equivalently, $r_s < 1$. In the leading order in $\kappa/ k_F$ one finds in 3D:
\begin{equation}
    \begin{split}
        U &= \frac{2\pi e^2}{9}\left( 3\ln \frac{1}{r_s} - 2 c^2 \frac{\tilde{\alpha}^2}{r_s^4}- 2 c^2 \frac{\tilde{\beta}^2}{r_s^4} \right);\\
        V &= \frac{2\pi e^2}{9} c \frac{\tilde{\alpha}}{r_s^2}\left(3-2c \frac{\tilde{\beta}}{r_s^2}\right);\quad 
        W = \frac{2\pi e^2}{45} c^2 \frac{\tilde{\alpha}^2}{r_s^4}\,,
    \end{split}
\end{equation}
where $\tilde{\alpha} = \alpha/(e a_B^2)$, $\tilde{\beta} = \beta/(e a_B^2)$, and  $c=(9\pi/4)^{2/3}\approx 3.7$. 

\begin{figure}[htb]
    \includegraphics[width=0.9\linewidth]{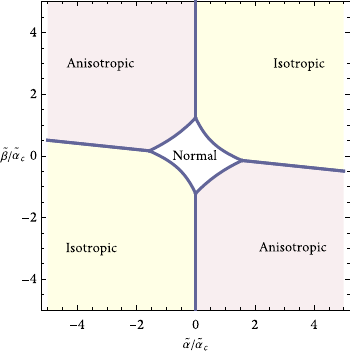}
    \caption{The diagram of p-wave superconducting instability according to the variational calculation. The isotropic phase corresponds to the singlet eigenvalue~\eqref{Eq:9x9_eigenvalues}, and the anisotropic to a quintet one. The normal region corresponds to all eigenvalues~\eqref{Eq:9x9_eigenvalues} being positive.  
 \label{Fig:diag_sup}}
\end{figure}

The absence of a term linear in $\tilde \beta$ in $U$ is a reflection of the fact that for the Coulomb potential the Darwin Hamiltonian~\eqref{Eq:Darwin} is $\sim \grad^2_{\bm{r}}(1/r) \sim \delta(\bm{r})$. It thus does not influence the energy of the $p-$wave state, which has zero anomalous expectation at coinciding spatial points. 

The eigenvalues in Eq.~\eqref{Eq:9x9_eigenvalues} become negative if the normalized couplings exceed a critical value, given by 
\begin{equation}
    \tilde{\alpha_c} = r_s^2 \frac{\sqrt{\ln \tfrac{1}{r_s}}}{c}\,.
\end{equation}
We notice that, since $\ln(1/r_s)>1 $, this critical value exceeds the bound~\eqref{Eq:saturation}, typical for the Kane-like multiband models.\footnote{We are grateful to V.~Kozii and J.~Ruhman for pointing out this observation.} It 
is possible, though, that other mechanisms of giant SOI may result in 
$\alpha >\alpha_c$. Particularly promising are graphene on TMDs and van der Waals materials with heavy adatoms. Verifying this possibility necessitates further investigation of fundamental physical mechanisms that give rise to an extremely strong Rashba effect.

The character of the superconducting instability depends on the sign and ratio of the coupling constants. This is illustrated by Fig.~\ref{Fig:diag_sup}, showing the emergence of distinct superconducting phases in 3D for various values of the coupling constants. For the sake of completeness, the diagram is shown in the extended parameter space; recall that in the Kane-like models the accessible region is restricted to $\tilde\beta>0$ and $-\tilde\beta/2 \le \tilde\alpha \le \tilde\beta$.

In the first and third quadrants of the phase diagram, Fig.~\ref{Fig:diag_sup}, the most unstable direction is associated with the singlet eigenvalue $U+4V$ of the $T^{\mu\mu',\nu\nu'}$ form. The corresponding eigenvector has equal components in $xx$, $yy$, and $zz$ directions. This describes the isotropic $p-$wave order parameter with $\bm{d}_{\bm{p}}\propto \bm{p}$. As shown in the next section, it leads to a node-less state with a gap isotropic around the Fermi surface. Such state is similar to the B phase of superfluid \isotope{He}{3}~\cite{wolfle}. In the second and fourth quadrants, the most unstable direction corresponds to the eigenvalue $U-2V+6W$. All five corresponding states break the rotational symmetry. One representative unstable direction is given by $\bm{d}_{\bm{p}}\propto (p^x, -p^y,0)$, which results in $p^x\pm ip^y$ order parameters for up/down spin components with nodes along the $p^z$-direction. This state is similar to the A phase of \isotope{He}{3}~\cite{wolfle}.

It is worth noticing that in 3D the sign of $\alpha$ has a physical meaning and, at fixed $\beta$, there is no invariance with respect to $\alpha\to -\alpha$ transformation, similar to quasi-2D systems with non-zero curvature~\cite{magarill1998ballistic}. Notice that in flat 2D systems the situation is different: in this case both $\nu$ and $\nu'$ are confined to the $x,y$ plane, and so are $\mu,\mu'$. This confines the quadratic form to be a $4\times 4$ matrix, $T^{\mu\mu',\nu\nu'} = U\delta_{\mu\nu}\delta_{\mu'\nu'}+2V \varepsilon^{\mu\nu z}\varepsilon^{\mu'\nu' z}$, with a symmetric spectrum of $U\pm 2V$, where $V$ is linear in $\alpha$. This reflects the fact that in 2D the combined $\alpha\to -\alpha$ and up $\to$ down transformations is the symmetry of the Rashba Hamiltonian.

In the opposite limit of strongly screened, short-ranged interactions where $r_s > 1$ and $\kappa^2/k_F^2 > 1$, the interaction constants are given by:
\begin{equation}
 U = C\tilde{\beta},\quad V = C\frac{\tilde{\alpha}}{2},\quad W = 0,
\end{equation}
where $C = \frac{4 \pi^2 \sqrt{c} e^2}{9 r_s} \left(1 - c \frac{\tilde{\beta}}{r_s^2}\right)$. The corresponding eigenvalues
\begin{equation}
 C \{2 \tilde{\alpha} + \tilde{\beta}, \tilde{\alpha} + \tilde{\beta}, -\tilde{\alpha} + \tilde{\beta}\}
\end{equation}
remain positive within the constraints imposed by Eq.~\eqref{Eq:inequalities} for the Kane-like models, provided that the Darwin coupling is less than critical, $\tilde{\beta} < r_s^2/c$. These findings, while consistent with previous studies~\cite{PhysRevB.96.214514,PhysRevX.9.031046}, do not necessarily preclude the possibility of superconductivity in a different parameter range.

\subsection{Self-consistency condition}

To figure out the fate of the $p-$wave instability, identified above, we formulate $4\times 4$ Bogoliubov-de-Gennes (BdG) Hamiltonian in the Nambu times spin space. In the basis where 
$\bar\Psi_{\bm{p}}=(\bar\psi_{\bm{p}\uparrow}, \bar\psi_{\bm{p}\downarrow}, 
\psi_{-\bm{p}\uparrow}, \psi_{-\bm{p}\downarrow})$, it takes the form:
\begin{equation}
\label{Eq:HamBCS-1}
    \hat{H}_{\mathrm{BdG}} =  \sum_{\bm{p}}\, \bar\Psi_{\bm{p}}
    \begin{pmatrix}
    \xi_{\bm{p}} \hat\sigma_0  & \hat\Delta_{\bm{p}} \\ 
    \hat\Delta_{\bm{p}}^\dagger  &  - \xi_{\bm{p}} \hat\sigma_0
    \end{pmatrix}    \Psi_{\bm{p}}, 
\end{equation} 
where $\hat\Delta_{\bm{p}}$ is a symmetric matrix in the spin space and an odd function of momentum, $\bm{p}$, which maybe conveniently 
parameterized by a complex vector, $\bm{\Delta}_{\bm{p}}$, cf.\ Eq.~\eqref{Eq:anomalous-1}, 
\begin{equation}
    \hat\Delta_{\bm{p}}   =
    (\bm{\Delta}_{\bm{p}}\!\cdot\! \bm{\hat{\sigma}})\, i\hat{\sigma}_y\,.  
\end{equation} 
The spectrum of the BdG Hamiltonian~\eqref{Eq:HamBCS-1} is determined by the secular equation $\det \left[(E_{\bm{p}}^2 - \xi_{\bm{p}}^2)\hat\sigma_0 - \hat\Delta_{\bm{p}} \hat\Delta_{\bm{p}}^\dagger \right]= 0$. One notices that 
\begin{equation}
    \label{Eq:Delta-square}
    \hat\Delta_{\bm{p}} \hat\Delta_{\bm{p}}^\dagger = \abs{\bm{\Delta}_{\bm{p}}}^2 \hat\sigma_0 - 
    i[\bar{\bm{\Delta}}_{\bm{p}}\times \bm{\Delta}_{\bm{p}}]\cdot \bm{\hat{\sigma}}\,.
\end{equation} 
We will limit ourselves to states where the vector $\bm{\Delta}_{\bm{p}}$ is real, up to an overall phase (the so-called unitary states~\cite{leggett2006quantum}). Under this assumption $[\bar{\bm{\Delta}}_{\bm{p}}\times \bm{\Delta}_{\bm{p}}]=0$ and thus the spectrum of the BdG Hamiltonian is degenerate and given by $\{E_{\bm{p}},E_{\bm{p}},-E_{\bm{p}},-E_{\bm{p}}\}$, where
\begin{equation}
    \label{Eq:BdG-energy}
    E_{\bm{p}} =\sqrt{\xi_{\bm{p}}^2 + \abs{\bm{\Delta}_{\bm{p}}}^2}. 
\end{equation} 
Therefore the  Hamiltonian~\eqref{Eq:HamBCS-1} maybe written as 
\begin{equation}
\label{Eq:HamBCS-3}
    \hat{H}_{\mathrm{BdG}} =  \sum_{\bm{p}}\,  E_{\bm{p}}  \, 
    \bar\Psi_{\bm{p}}U^\dagger_{\bm{p}}\,  (\hat\sigma_0\hat\tau_z)\, U_{\bm{p}} \Psi_{\bm{p}}\,, 
\end{equation} 
where $\tau_z$ is the Pauli matrix in the Nambu space and $U_{\bm{p}}$ is a $4\times 4$ unitary transformation, which diagonalizes the matrix. One thus finds standard equal time equilibrium expectation values for the unitary rotated fermions 
\begin{equation} 
    U_{\bm{p}} \av*{\Psi_{\bm{p}} \bar\Psi_{\bm{p}}} U^\dagger_{\bm{p}} = 
    \begin{pmatrix}
    (1-n_{\bm{p}})\hat\sigma_0  & 0 \\ 
    0  &  \!\!n_{\bm{p}}\hat\sigma_0
    \end{pmatrix}= 
    \frac{\hat\openone}{2}  + \frac{1-2n_{\bm{p}}}{2}\hat\sigma_0\hat\tau_z\,, 
\end{equation}
where $n_{\bm{p}}= n(E_{\bm{p}})$ is the Fermi function.  Therefore the BdG expectation is given, cf.\ Eqs.~\eqref{Eq:HamBCS-1} and~\eqref{Eq:HamBCS-3},  
\begin{equation}
\label{Eq:Bdg-expectation}
    \av*{\Psi_{\bm{p}} \bar\Psi_{\bm{p}}}  = \frac{\hat\openone}{2} + 
    \frac{1-2n_{\bm{p}}}{2E_{\bm{p}}}  
    \begin{pmatrix}
    \xi_{\bm{p}} \hat\sigma_0  & \hat\Delta_{\bm{p}} \\ 
    \hat\Delta_{\bm{p}}^\dagger  &  - \xi_{\bm{p}} \hat\sigma_0
    \end{pmatrix}\,.
\end{equation} 
Comparing this expression with the definition of the anomalous average~\eqref{Eq:anomalous-1}, one finds that 
\begin{equation}
    \label{Eq:anomalous-3}
 \hat{d}^{\,\intercal}_{\bm{p}}=  -  \frac{\tanh E_{\bm{p}}/2T }{2E_{\bm{p}}}   \,\, \hat\Delta_{\bm{p}}\,. 
\end{equation} 

To close the self-consistency loop one compares the BdG Hamiltonian~\eqref{Eq:HamBCS-1} with the original BCS Hamiltonian~\eqref{Eq:HamBCS} and notices that 
\begin{equation}
    \hat \Delta_{\bm{p}} =  \sum_{\bm{p}'} \,  \mathcal{U}_{\bm{p}-\bm{p}'} \hat\Gamma_{\bm{p}\bm{p}'}\, \hat{d}^{\,\intercal}_{\bm{p}'} \,\hat\Gamma^{\intercal}_{\bm{p}\bm{p}'}\,.  
\end{equation}
Combining this with Eq.~\eqref{Eq:anomalous-3}, one obtains the self-consistency equation for the matrix order parameter
\begin{equation}
                \label{Eq:self-consistency}
    \hat \Delta_{\bm{p}} = - \sum_{\bm{p}'} \,  \mathcal{U}_{\bm{p}-\bm{p}'} 
    \frac{\tanh(E_{\bm{p}'}/2T)}{2E_{\bm{p}'}}\, \,  
    \hat\Gamma_{\bm{p}\bm{p}'}\, \hat\Delta_{\bm{p}'} \,\hat\Gamma^{\intercal}_{\bm{p}\bm{p}'}\,. 
\end{equation}
Using Eq.~\eqref{Eq:VertexMomentum} for the vertex one finds
\begin{widetext}
    \begin{equation}
        \label{Eq:self-consistency-1}
        \bm{\Delta}_{\bm{p}} = -\sum_{\bm{p}'} \mathcal{U}_{\bm{p}-\bm{p}'} 
        \frac{\tanh(E_{\bm{p}'}/2T)}{2E_{\bm{p}'}}\left\{ (\breve{e}^2-\alpha^2 \mathcal{M}^2)\bm{\Delta}_{\bm{p}'}+2\breve{e}\alpha [\bm{\Delta}_{\bm{p}'}\times \bm{\mathcal{M}}] + 2\alpha^2(\bm{\Delta}_{\bm{p}'}\cdot \bm{\mathcal{M}})\bm{\mathcal{M}}\right\}  \,.
    \end{equation}
\end{widetext}

For the isotropic state with $\bm{\Delta}_{\bm{p}} = \bm{p}\,{\Delta}_{{p}}/p$, the BdG energies are also isotropic, given by $E_{{p}} =\sqrt{\xi_{{p}}^2 + \abs{{\Delta}_{{p}}}^2}$. Performing the angular integration, one observes that the isotropic ansatz indeed satisfies the self-consistency equation, provided that the amplitude of the order parameter,  ${\Delta}_{{p}}\approx {\Delta}_{{k_F}}=\Delta$, is a solution of 
\begin{equation}
    \label{Eq:self-consistency-3}
    1 = - \frac{3\nu_3}{4 k_F^2}(U+4V) \int d\xi_p\, \frac{\tanh \frac{\sqrt{\xi_p^2 +\Delta^2}}{2T}}{\sqrt{\xi_p^2 +\Delta^2}}\,, 
\end{equation}
where we anticipated that the $\xi_p$ integration is limited to a narrow vicinity of the Fermi energy, allowing us to put $p\approx p'\approx k_F$. The self-consistency equation has a solution provided that $U+4V<0$, which is exactly the instability condition, obtained in the previous section for the isotropic state. The region of existence of the isotropic p-wave order is thus the same as in Fig.~\ref{Fig:diag_sup}. The corresponding critical temperature is given by
\begin{equation}
    T_c \approx \Lambda\exp\left\{ - \frac{2 k_F^2}{3\nu_3 \abs*{U+4V}} \right\},
\end{equation}
where $\Lambda$ is an energy scale over which the SOI ``constant'', $\alpha$, decreases. This is the  Balian-Werthamer phase~\cite{PhysRev.131.1553}, observed as the B phase of \isotope{He}{3}. 

For the anisotropic case the linear analysis indicates five degenerate unstable directions. They correspond to states with $\bm{\Delta}_{\bm{p}} \sim (p^x,-p^y,0)$, or $\bm{\Delta}_{\bm{p}} \sim (p^y,p^x,0)$ and those obtained by permutations of the three axis. 
The anisotropic order parameter, while breaking spontaneously the global $U(1)$ and rotational symmetries, does respect the time reversal symmetry, since the spin up condensate maps onto spin down one upon time reversal transformation.
This is the analog of Anderson-Brinkman-Morel~\cite{PhysRev.123.1911} superconductor, observed as a phase A of \isotope{He}{3}. Although the full analysis of the self-consistency condition in this case is rather involved, it is likely that such states  exhibit a nodal order parameter with the two nodes along, say,  $p^z$-axis. 
The corresponding quasiparticle spectrum is also anisotropic, schematically of the form  $E_{\bm{p}} =\sqrt{\xi_{{p}}^2 + \abs{{\Delta}_{{p}}}^2\sin^2 \theta_{\bm{p}}}$, though its precise angular dependence is more complicated. One can not exclude, though, a possibility that a certain (non-unitary) combination of the five unstable directions results in an anisotropic nodeless state.  

In 2D, the amplitude of the order parameter is a solution of 
\begin{equation}
    \label{Eq:self-consistency-2d}
    1 = - \frac{\nu_2}{2 k_F^2}(U\pm 2V) \int d\xi_p\, \frac{\tanh \frac{\sqrt{\xi_p^2 +\Delta^2}}{2T}}{\sqrt{\xi_p^2 +\Delta^2}}\,, 
\end{equation}
where $\pm$ sign corresponds to $\bm{\Delta}_{\bm{p}} \sim (p^x,\pm p^y,0)$, which correspond to $\alpha\to -\alpha$ transformation. Both of these states are isotropic in 2D. They correspond to $p_x\pm ip_y$ orbital order for spin up and $p_x\mp ip_y$ order for spin down. Since this state is gapped and belongs to $DIII$ symmetry class, it admits  $Z_2$ topological index in 2D~\cite{haim2019time}. The criterion for such a state to exist is $|V| > U/2$, where the corresponding 2D constants are given by

\begin{equation}
    \begin{split}
        U &= 2e^2 k_F\left( \ln \frac{1}{r_s} +\frac83 \frac{\tilde{\beta}}{r_s^2} -\frac{16}{15}\frac{\tilde{\alpha}^2}{r_s^4}- \frac{32}{5} \frac{\tilde{\beta}^2}{r_s^4} \right);\\
        V &= e^2 k_F\frac{16 \tilde{\alpha}}{15r_s^2}\left( 5- 8\frac{\tilde{\beta}}{r_s^2}\right)\,.
    \end{split}
\end{equation}

The time reversal symmetry may be also broken, if  a certain spin polarization spontaneously develops. This gives rise to a superconducting phase similar to the A\textsubscript{1} phase in \isotope{He}{3}. 
Since the two spin components are essentially independent, like in a normal Fermi liquid, the spin susceptibility is expected to be close to its normal state value. However, as explained in the next section, it is actually enhanced, which may lead to a ferromagnetic instability. 

\section{Ferromagnetism}
\label{Sec:TPT}
Here, we show that PSOI-mediated exchange interaction may lead to a ferromagnetism. Due to the spin-orbit coupling, it breaks rotational symmetry both in the spin and momentum spaces, resulting in an anisotropic Fermi surface. For a sufficiently large SOI constant, $\alpha$, the Fermi surface of the ferromagnet undergoes the topological Lifshitz transition from an ellipsoidal to a toroidal shape.  

\begin{figure}[htb]
    \includegraphics[width=0.5\linewidth]{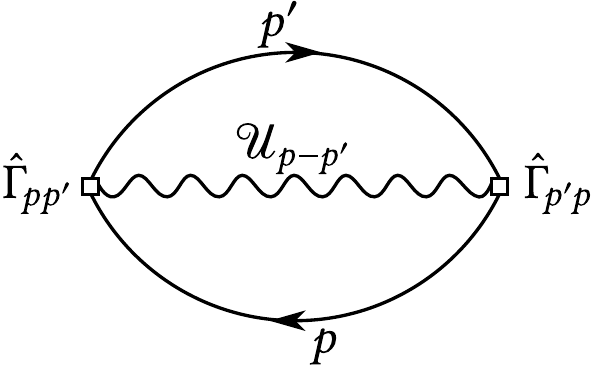}
    \caption{The first order diagram for the exchange energy of the electron gas. The momentum-dependent vertices $\hat{\Gamma}_{\bm{p}\bm{p}'}$ of Eq.~\eqref{Eq:VertexMomentum} reflect the PSOI contribution to the energy.\label{Fig:Fock}}
\end{figure}

\subsection{Mean-field theory}

We  treat the e-e interaction of Eq.~\eqref{Eq:HamInt} in a Hartree-Fock (HF) approximation. The Hartree (direct) contribution is not modified by the anomalous contributions. Indeed, for the Hartree part of Eq.~\eqref{Eq:HamInt} $\bm{p}_1=\bm{p}_2$ and $\bm{p}_1'=\bm{p}_2'$, nullifying the anomalous parts of interaction vertexes, $\hat\Gamma$, Eq.~\eqref{Eq:VertexMomentum}. As a result, the Hartree term is compensated by the positive charge of the lattice and is absent in  charge neutral systems. The Fock (exchange) contribution comes from the ``oyster'' diagram of Fig.~\ref{Fig:Fock}. It is given by
\begin{equation}
    \label{Eq:energy-trace}
    E_{\mathrm{xc}} = - \frac{1}{2} \sum_{\bm{p} \bm{p}'} \mathcal{U}_{\bm{p}-\bm{p}'} \Tr\{\hat{G}_{\bm{p}}\hat{\Gamma}_{\bm{p}\bm{p}'}\hat{G}_{\bm{p}'}\hat{\Gamma}_{\bm{p}'\bm{p}}\}\,,
\end{equation}
where $\hat{G}_{\bm{p}}$ is the equal-time electron Green function.

To reach a spin-polarized state, one may assume a symmetry-breaking in any given direction, say along the $z$-axis. Under this assumption, the Green function takes the form~\footnote{In principle, the Green function does not have to be diagonal for all values of momenta in the same basis. For example, one may think of the Green functions with the off-diagonal elements, e.g., of the form $\propto(\sigma_x p_x +\sigma_y p_y)$, signifying an interaction-induced single-particle RSOI\@. One may show, however, that such states have higher energy, than that given by the diagonal ansatz.}  
\begin{equation}
    \hat{G}_{\bm{p}} = \mathcal{G}_0(\bm{p}) \hat{\sigma}_0 + \mathcal{G}_3(\bm{p}) \hat{\sigma}_3\;,
\end{equation}
or explicitly,
\begin{equation}
    \hat{G}_{\bm{p}} =
                        \begin{pmatrix}
                            n_{+}(\bm{p}) & 0\\
                            0 & n_{-}(\bm{p})
                        \end{pmatrix}\,,
\end{equation}
where $n_\pm (\bm{p})$ are occupation numbers of a state $\bm{p}$ by electrons with the two spin projections.   
Computing the trace in Eq.~\eqref{Eq:energy-trace} one obtains
\begin{widetext}
    \begin{equation}
        \label{Eq:Exchange-energy}
        \begin{split}
            E_{\mathrm{xc}} = &{}- \sum_{\bm{p} \bm{p}'} \breve{e}^2 \mathcal{U}_{\bm{p}-\bm{p}'} \left\{\mathcal{G}_0(\bm{p}) \mathcal{G}_0(\bm{p}') + \mathcal{G}_3(\bm{p}) \mathcal{G}_3(\bm{p}')\right\}\\
            &{}- \sum_{\bm{p} \bm{p}'} \alpha_{p p'}^2 \mathcal{U}_{\bm{p}-\bm{p}'} \left\{\mathcal{G}_0(\bm{p}) \mathcal{G}_0(\bm{p}')(\bm{\mathcal{M}}\cdot \bm{\mathcal{M}}) + \mathcal{G}_3(\bm{p})\mathcal{G}_3(\bm{p}')(\mathcal{M}_z^2-\mathcal{M}_x^2-\mathcal{M}_y^2)\right\}\,.
        \end{split}
    \end{equation}
\end{widetext}
The first line of this equation represents the familiar exchange energy of the Coulomb-interacting electron gas~\cite{Vignalebook}. The second line captures the contribution of the PSOI to the exchange. Unlike in the Cooper channel discussed above, the PSOI contribution here is entirely decoupled from the Darwin interaction. Both ultimately reduce the total energy, potentially driving the system toward Bloch ferromagnetism. The Darwin correction introduces a rotationally invariant modification to the electron dispersion, similarly to the normal Coulomb potential. In stark contrast, the PSOI contribution yields a far more intriguing behavior. It can lead to a strongly anisotropic electron dispersion, resulting in spontaneous rotational symmetry breaking and topological Lifshitz transitions. In this section, we will focus exclusively on the effects of PSOI\@. Our findings should remain qualitatively consistent in more general models that include anomalous corrections to scalar potential.

\subsection{Paramagnetic phase}
\label{Sec:Para}

In the paramagnetic electron state $n_{-}(\bm{p})=n_{+}(\bm{p}) \equiv n(\bm{p})$ and Eq.~\eqref{Eq:Exchange-energy}  takes the form
\begin{equation}
    \label{Eq:Energy-nonpolarized}
    \begin{split}
        E_{\mathrm{xc}} = &{}- \sum_{\bm{p} \bm{p}'} e^2 \mathcal{U}_{\bm{p}-\bm{p}'} n(\bm{p}) n(\bm{p}')\\
        &{}- \sum_{\bm{p} \bm{p}'} \alpha_{p p'}^2 \mathcal{U}_{\bm{p}-\bm{p}'} n(\bm{p}) n(\bm{p}') \mathcal{M}^2 \,.
    \end{split}
\end{equation}
Taking variation with respect to $n(\bm{p})$, one finds  the HF effective dispersion of the form 
\begin{equation}
    \label{Eq:MF-nonpolarized}
    \begin{split}
        V_{\bm{p}}^{\mathrm{HF}} = {}&  \frac{ p^2}{2m} - 2 \sum_{\bm{p}'} e^2 \mathcal{U}_{\bm{p}-\bm{p}'}  n(\bm{p}')\\
        &{}- 2 \sum_{\bm{p}'} \alpha_{p p'}^2 \mathcal{U}_{\bm{p}-\bm{p}'} n(\bm{p}') \mathcal{M}^2 \,.
    \end{split}
\end{equation}

To keep things analytically tractable, we will temporarily disregard the momentum dependence of the interaction, $\mathcal{U}_{\bm{p}-\bm{p}'} \equiv \mathcal{U}_0$, and assume a specific momentum dependence for the SOI magnitude, reflecting its decrease with increasing momentum
\begin{equation}
    \alpha_{p p'}^2 = \alpha^2 \frac{\Lambda^2}{(\Lambda + p^2)(\Lambda + {p'}^2)}\,,
\end{equation}
where $\Lambda/2m$ is an energy scale of the order of the band gap. 

These model restrictions will be subsequently relaxed to account for realistic long-ranged interactions with dielectric screening and the precise microscopic momentum-dependence of the SOI\@. A numerical analysis of 2D Rashba systems that includes these factors is presented in Appendix~\ref{Sec:Ferro-2D}.

The ground state is constructed by filling the quantum states of the lowest energy. These states have the least momentum when the HF potential $ V_{\bm{p}}^{\mathrm{HF}} $ increases monotonically with $ p $. However, this is not the case with the potential of Eq.~\eqref{Eq:MF-nonpolarized}. The momentum dependence of $ \bm{\mathcal{M}}(\bm{p},\bm{p}') $ makes the effective dispersion non-monotonous, creating a moat-like dispersion. Consequently, it becomes beneficial for electrons to populate states with higher momentum — a distinct feature of the PSOI\@. This fact eventually leads to the topological Lifshitz  transition.

To see this, first assume that electrons in a 3D gas with concentration $n$ form a Fermi sea enclosed by a spherical Fermi surface of radius $k_F = {(3 \pi^2 n)}^{1/3}$, with $n(\bm{p}) = \theta(k_F - \abs{\bm{p}})$. Such distribution function generates a spherically symmetric effective dispersion of the form
\begin{equation}
    V_{\bm{p}}^{\mathrm{HF}} = \frac{ p^2}{2m} - e^2 \mathcal{U}_0 n - \frac{2\alpha^2}{3 \pi^2} \mathcal{U}_0  \frac{\Lambda^2 p^2}{\Lambda + p^2}\, \Xi(k_F)\,.    
\end{equation}
The factor $\Xi(k_F)$ is given by
\begin{equation}
    \label{Eq:Xi-function}
    \Xi(k_F) = \frac{k_F^3}{3} - k_F \Lambda + \Lambda^{\frac32} \arctan \frac{k_F}{\sqrt{\Lambda}} \,,
\end{equation}
and interpolates between  $\Xi(k_F) \approx k_F^5/5\Lambda$ for $k_F \ll \sqrt{\Lambda}$ and $\Xi(k_F) \approx k_F^3/3$ for $k_F \gg \sqrt{\Lambda}$.

\begin{figure}[htb]
	\includegraphics[width=0.9\linewidth]{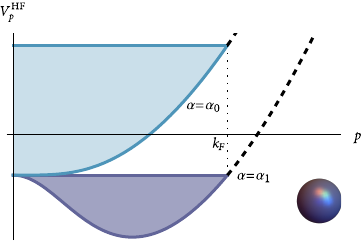}
	\caption{The effective dispersion $V_{\bm{p}}^{\mathrm{HF}}$ as a function of momentum for two values of the SOI magnitude. The position of the Fermi momentum is shown by a vertical dotted line. The occupied states are shown in color. The Fermi surface is a sphere, as shown in the inset.\label{Fig:mf-para-trivial}}
\end{figure}

The effective mass of the electron diverges when the SOI magnitude $\alpha$ equals to
\begin{equation}
    \alpha_0 = \sqrt{\frac{3 \pi^2 }{4 m \mathcal{U}_0 \Lambda \Xi(k_F)}}\,,
\end{equation}
with a flat band forming around the Brillouin zone center. The dependence of the effective dispersion on the momentum becomes non-monotonous as soon as $\alpha > \alpha_0$, see Fig.~\ref{Fig:mf-para-trivial}. A minimum appears for a finite momentum, which depends on $\alpha$. For a given electron concentration, there exist a critical $\alpha_1$ such that for $\alpha \le \alpha_1$ all states with $\abs{\bm{p}} \le k_F$ are occupied. This renders the Fermi surface as a sphere, in agreement with our initial assumption. 

\begin{figure}[htb]
	\includegraphics[width=0.9\linewidth]{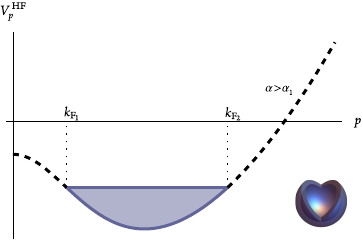}
	\caption{The effective dispersion $V_{\bm{p}}^{\mathrm{HF}}$ as a function of momentum for the larger-than-critical value of the SOI magnitude. The populated states are enclosed by a shell-like Fermi surface.\label{Fig:mf-para-top}}
\end{figure}

The critical value of $\alpha_1$ is inferred from 
\begin{equation}
    V_{\bm{p}}^{\mathrm{HF}} \big|_{p=0} = V_{\bm{p}}^{\mathrm{HF}} \big|_{p=k_F}
\end{equation}
to give $ \alpha_1 = \alpha_0 \sqrt{1+ k_F^2/\Lambda}$. 
For $ \alpha > \alpha_1 $ states with the lowest energy no longer correspond to the smallest momenta. Instead, they are found in a range of momenta  $k_{F_1} \le \abs{\bm{p}} \le k_{F_2}$, forming a hollow spherical shell. This behavior is illustrated in Fig.~\ref{Fig:mf-para-top}. Such a transformation signals a Lifshitz  transition, changing a topology of the Fermi surface.  

The corresponding distribution function takes the form 
\begin{equation}
    \label{Eq:shell-distribution}
    n(\bm{p}) =\theta(\abs{\bm{p}} - k_{F_1})\theta(k_{F_2} - \abs{\bm{p}})\,,
\end{equation}
where the two Fermi momenta are  related by a requirement of the fixed electron concentration, $k_{F_2}^3 - k_{F_1}^3 = 3\pi^2 n$. Such a distribution gives rise to a self-consistent solution provided that
\begin{equation}
    \label{Eq:MF-nonpolarized-scc}
    V_{\bm{p}}^{\mathrm{HF}} \big|_{p=k_{F_1}} = V_{\bm{p}}^{\mathrm{HF}} \big|_{p=k_{F_2}}\,.
\end{equation}
The effective dispersion  calculated with the distribution function~\eqref{Eq:shell-distribution} is 
\begin{equation}
    \label{Eq:MF-shell}
    V_{\bm{p}}^{\mathrm{HF}} = \frac{ p^2}{2m} - e^2 \mathcal{U}_0 n - \frac{2\alpha^2 \mathcal{U}_0}{3 \pi^2} \frac{\Lambda^2 p^2}{\Lambda + p^2}\left[\Xi(k_{F_2}) - \Xi(k_{F_1})\right]\,.    
\end{equation}

Equations~\eqref{Eq:MF-nonpolarized-scc}-\eqref{Eq:MF-shell} lead to the following relation that allows one to find the two Fermi momenta,
\begin{equation}
    \alpha^2[\Xi(k_{F_2}) - \Xi(k_{F_1})] = \left(1+ \frac{k_{F_1}^2}{\Lambda}\right) \left(1+ \frac{k_{F_2}^2}{\Lambda}\right) \frac{3 \pi^2 }{4 m \mathcal{U}_0 \Lambda}\,.
\end{equation}
As soon as $\alpha > \alpha_1$, the solution of this equation emerges with $k_{F_1}>0$, which determines the position of the shell boundaries in the momentum space.

The total energy per one electron, calculated self-consistently from  Eq.~\eqref{Eq:Energy-nonpolarized} and taking into account the kinetic energy of the electrons filling the shell in the momentum space, equals
\begin{equation}
    \begin{split}
        \varepsilon =&{} \frac{1}{10 \pi^2 n}\frac{k_{F_2}^5 - k_{F_1}^5}{m} -e^2 \mathcal{U}_0 \frac{n}{4}\\
        &{} - \frac{\alpha^2}{6 \pi^4 n}\mathcal{U}_0 \Lambda^2 {\left[\Xi(k_{F_2}) - \Xi(k_{F_1})\right]}^2\,.
    \end{split}
\end{equation}
This expression is valid for both shell-like ($\alpha > \alpha_1$) and spherical ($\alpha \le \alpha_1$) Fermi surfaces, with $k_{F_1} = 0$ in the latter case.

\subsection{Ferromagnetic phase}
\label{Sec:Ferro}

In a fully spin-polarized case\footnote{It is generally believed that in 2D systems the ferromagnetic transition is of the first order (Bloch transition), whereas a second-order transition (Stoner) is not realized under the assumption of spin-independent interactions. How PSOI affects this scenario remains an area of ongoing research. In 3D systems, a second-order transition is possible, but our focus here is on demonstrating the instability with respect to full magnetization, rather than exploring transitional states with partial magnetization.} of $n_{-}(\bm{p})=0$, the Eq.~\eqref{Eq:Exchange-energy} reduces to
\begin{equation}
    \label{Eq:Energy-polarized}
    \begin{split}
        E_{\mathrm{xc}} = &{}- \frac{1}{2} \sum_{\bm{p} \bm{p}'} e^2 \mathcal{U}_{\bm{p}-\bm{p}'} n(\bm{p}) n(\bm{p}')\\
        &{}-  \frac{1}{2} \sum_{\bm{p} \bm{p}'} \alpha_{p p'}^2 \mathcal{U}_{\bm{p}-\bm{p}'} n(\bm{p}) n(\bm{p}') \mathcal{M}_z^2 \,,
    \end{split}
\end{equation}
with the corresponding HF dispersion given by
\begin{equation}
    \label{Eq:MF-polarized}
    \begin{split}
        V_{\bm{p}}^{\mathrm{HF}} = {}&  \frac{ p^2}{2m} -  \sum_{\bm{p}'} e^2 \mathcal{U}_{\bm{p}-\bm{p}'}  n(\bm{p}')\\
        &{}- \sum_{\bm{p}'} \alpha_{p p'}^2 \mathcal{U}_{\bm{p}-\bm{p}'} n(\bm{p}') \mathcal{M}_z^2 \,.
    \end{split}
\end{equation}

\begin{figure}[htb]
	\includegraphics[width=0.8\linewidth]{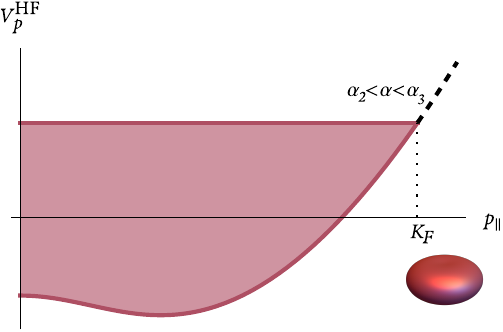}
	\caption{The effective dispersion $V_{\bm{p}}^{\mathrm{HF}}$ as a function of the in-plane momentum $p_{\parallel}$ at $p_z=0$ for weak SOI\@. \label{Fig:mf-ferro-trivial}}
\end{figure}

Assuming for a moment a fully spin-polarized spherical Fermi surface  with
\begin{equation}
\label{Eq:Fermi_momentum_FM}
    K_F = \sqrt[3]{2} k_F = {(6 \pi^2 n)}^{\frac13},
\end{equation}
results in an anisotropic HF effective dispersion 
\begin{equation}
    \label{Eq:MF-polarized-1}
    V_{\bm{p}}^{\mathrm{HF}} = \frac{ p^2}{2m} - e^2 \mathcal{U}_0 n - \frac{\alpha^2 \mathcal{U}_0}{6 \pi^2}  \frac{\Lambda^2 p^2_{\parallel}}{\Lambda + p^2}\, \Xi(K_F)\,,
\end{equation}
where $\bm{p}_{\parallel} \equiv (p_x,p_y,0)$ is the in-plane momentum. This indicates an increased in-plane effective mass and therefore an anisotropic pancake shape of the Fermi surface, see Fig. \ref{Fig:mf-ferro-trivial}.

\begin{figure}[htb]
	\includegraphics[width=0.9\linewidth]{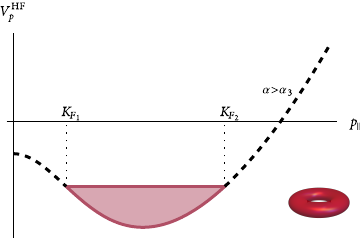}
	\caption{The effective dispersion $V_{\bm{p}}^{\mathrm{HF}}$ as a function of the in-plane momentum $p_{\parallel}$ at $p_z=0$, for the larger-than-critical value of the SOI magnitude. The Fermi surface is shaped like a torus.\label{Fig:mf-ferro-top}}
\end{figure}

According to Eq.~\eqref{Eq:MF-polarized-1}, surfaces of a constant energy may be parameterized by
\begin{equation}
\label{Eq:Oblate_FS}
    (p^2 + \Lambda)(p^2 + \zeta) = 4 K^2 p_{\parallel}^2\,,
\end{equation}
with some constants $\zeta$ and $K$ to be determined self-consistently. For a small $\alpha$ the Fermi surface has a flattened ellipsoidal shape. However, upon the increase of $\alpha$ it undergoes a qualitative change. The in-plane effective mass diverges for the SOI magnitude of
\begin{equation}
    \alpha_2 = \sqrt{\frac{3 \pi^2 }{m \mathcal{U}_0 \Lambda \Xi(K_F)}}\,.
\end{equation}
For $\alpha>\alpha_3$, where  $\alpha_3 = \alpha_2 \sqrt{1+ K_F^2/\Lambda}$, the Fermi surface acquires a toroidal shape, with the in-plane momenta limited to a range $K_{F_1} \le p_{\parallel} \le K_{F_2}$, while the normal momenta are concentrating near the center, $\abs{p_z} < K_{F_z}$. As a result, the system undergoes the topological Lifshitz transition, with the Fermi surface genus changing from zero to one, see Fig. \ref{Fig:mf-ferro-top}.

One may show (see Appendix~\ref{Sec:Torus}) that the Fermi surface of Eq.~\eqref{Eq:Oblate_FS} provides a self-consistent solution to Eq.~\eqref{Eq:MF-polarized}. The total energy per one electron in the toroidal phase is equal to
\begin{equation}
    \varepsilon = \frac{1}{4 m }\left( 4 K^2 - \Lambda - \zeta \right) -e^2 \mathcal{U}_0 \frac{n}{2}
        - \frac{\alpha^2 \mathcal{U}_0 \Lambda^2 \Xi^2}{4 n}\,,
\end{equation}
with parameters $K$, $\zeta$ and $\Xi$ detailed in the Appendix~\ref{Sec:Torus}. The results for the elliptical phase are quite similar; however, we have omitted them for brevity.

\subsection{Ferromagnetic Transitions}
\label{Sec:FMT}

\begin{figure}[htb]
	\includegraphics[width=0.9\linewidth]{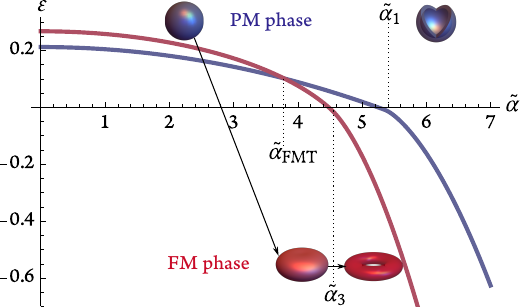}
	\caption{The energy per particle of a 3D electron gas, in units of $K_F^2/2m$, as a function of the dimensionless SOI magnitude $\tilde{\alpha}= \alpha K_F^2/e$, for $\tilde{\mathcal{U}} = 1$ and $\Lambda/K_F^2 = 1$. The red curve corresponds to the ferromagnetic (FM) phase, and the blue curve to the paramagnetic (PM) one. The Bloch transition occurs at $\tilde{\alpha} = \tilde{\alpha}_{\mathrm{FMT}}$.\label{Fig:phases-u1}}
\end{figure}

Figure~\ref{Fig:phases-u1} depicts the total HF energy per electron for both the paramagnetic and spin-polarized phases, plotted against the magnitude of the PSOI\@. This is represented for $ \tilde{\mathcal{U}} = 1 $, where the e-e interaction parameter is normalized by the density of states as:
\begin{equation}
    \tilde{\mathcal{U}} = e^2 \mathcal{U}_0 \frac{m K_F}{2\pi^2 \hbar^2}\,.
\end{equation}
At $\tilde{\alpha}=0$ the system is in the paramagnetic phase characterized by a spherical Fermi surface. As PSOI increases, the first order ferromagnetic transition occurs at a critical $\tilde{\alpha} = \tilde{\alpha}_{\mathrm{FMT}}$, which is accompanied by a Lifshitz transition to a state with an elliptical Fermi surface. In other words, both spin symmetry and the Fermi-surface symmetry simultaneously undergo a change at this point. As PSOI continues to increase, it induces a Lifshitz transition to a toroidal spin-polarized phase at $\tilde{\alpha} = \tilde{\alpha}_3$.

\begin{figure}[htb]
	\includegraphics[width=0.9\linewidth]{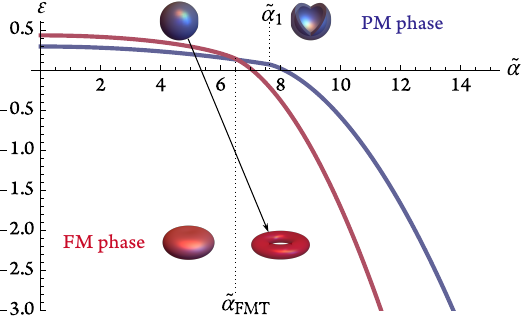}
	\caption{The energy per particle of a 3D electron gas as a function of the dimensionless SOI magnitude $\tilde{\alpha}= \alpha K_F^2/e$ for $\tilde{\mathcal{U}} = 0.5$. The red curve corresponds to the ferromagnetic phase, and the blue curve to the paramagnetic one.\label{Fig:phases-u05}}
\end{figure}

\begin{figure}[htb]
	\includegraphics[width=0.9\linewidth]{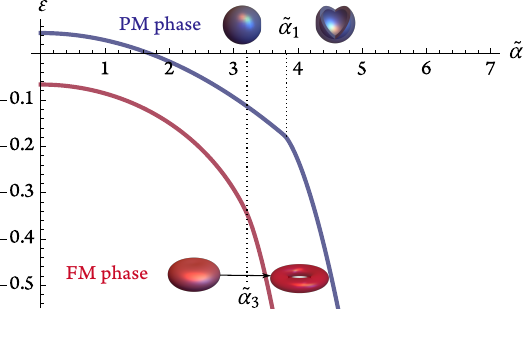}
	\caption{The energy per particle of a 3D electron gas as a function of the dimensionless SOI magnitude $\tilde{\alpha}= \alpha K_F^2/e$ for $\tilde{\mathcal{U}} = 2$. The red curve corresponds to the ferromagnetic phase, and the blue curve to the paramagnetic one.\label{Fig:phases-u2}}
\end{figure}

For a smaller interaction parameter of $\tilde{\mathcal{U}}=0.5$, the system can transition directly from the spherical paramagnetic phase to the toroidal spin-polarized phase, bypassing the elliptical phase, as depicted in Fig.~\ref{Fig:phases-u05}. Conversely, with increased interaction, the system defaults to a ferromagnetic state at zero SOI due to the conventional Bloch instability. Yet, any non-zero value of $\alpha$ ensues breaking of the Fermi surface's spherical symmetry. A further increase of $\alpha$ results into the Lifshitz transition from the ellipsoidal to the toroidal phase at $\tilde{\alpha} = \tilde{\alpha}_3$, as visualized in Fig.~\ref{Fig:phases-u2}.

\begin{figure}[htb]
	\includegraphics[width=0.9\linewidth]{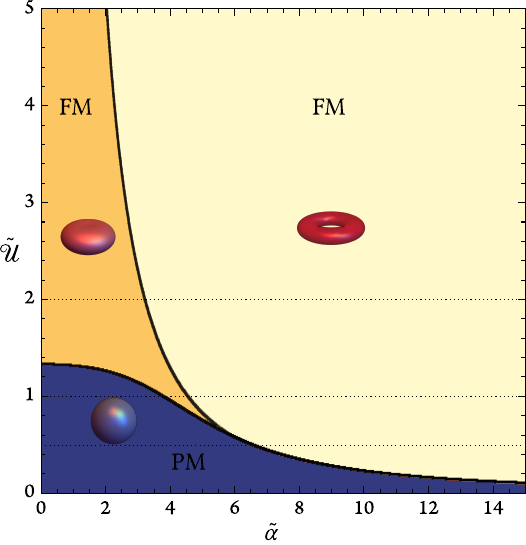}
	\caption{The phase diagram of the 3D Rashba system. Three horizontal cross-sections correspond to Figs.~\ref{Fig:phases-u1}--\ref{Fig:phases-u2}.\label{Fig:diag3D}}
\end{figure}

These findings are summarized in Fig.~\ref{Fig:diag3D}, showing the phase diagram in the coordinates of dimensionless SOI coupling $\tilde{\alpha}$ vs.\ the interaction strength $\tilde{\mathcal{U}}$. The  paramagnet phase exhibits the first order transition to the fully polarized ferromagnet. The latter may come with either flat ellipsoidal or toroidal Fermi surfaces with the topological Lifshitz transition between them. The hollow spherical shell paramagnetic phase, discussed above, is not realized (at least within the present model), since the transition to the FM phase takes place before it becomes favorable. It is worth mentioning  that the critical values of the parameters for the Bloch and Lifshitz transitions, obtained from the HF energy computations, are not expected to be quantitatively accurate. Correlation effects renormalize them, as indicated by the Monte Carlo simulations. Nevertheless, one may expect that the qualitative features of the phase diagram are still present, even with the account of the correlations. 

\subsection{Ferromagnetic Superconductors}

We conclude this section by a qualitative discussion of the possible coexistence of ferromagnetism and superconductivity in Rashba systems. Both ellipsoidal and toroidal fully polarized ferromagnets are susceptible to the development of the $p-$wave superconducting order, discussed in Section~\ref{sec:p-wave}. In this case only $\Delta^+_{\bm{p}}$ order parameter is realized, while $\Delta^0_{\bm{p}}$ and $\Delta^{-}_{\bm{p}}$ are absent (here~$\pm,0$ stay for $m=\pm 1$ and $m=0$ projection of the Cooper pair spin angular momentum). The toroidal ferromagnetic phase requires a critical SOI strength, $\tilde{\alpha}_F \sim r_s$, and for a dense gas where $\tilde{\alpha}_F \gg \alpha_c$, it formally falls within the parameter region that favors a superconducting ground state. In reality, if the disorder is sufficiently strong, the toroidal FM may not develop superconductivity down to zero temperature. 

The coexisting FM and superconducting orders break rotational, time-reversal and global $U(1)$ symmetries. This puts the model into symmetry class D. The latter is not topological in 3D, though it does admit an integer Chern number in 2D.  It is interesting to notice that the topological Lifshitz transition from elliptical to the toroidal Fermi surface has direct consequences for the superconducting order. Indeed, the $p-$wave superconductor with the order parameter $\Delta^+$ and the ellipsoidal  Fermi surface is  nodal with the two nodes on the flat parts of the ellipsoid. Once the Fermi surface transitions to the toroidal shape, the nodes disappear and the $p-$wave state  $\Delta^+$ is fully gapped (similarly to its 2D counterpart). This may lead to a weak first order transition between the nodal and gapped ground states, but its details require a further investigation.   

\subsection{Manifestations of the ferromagnetic transitions}

The FM state is, of course, associated with a net magnetic moment. This may be hard to detect due to, e.g., a domain structure (in presence of some uniaxial magnetic anisotropy). However, in the presence of strong SOI it should give rise to the anomalous Hall effect. Below we briefly discuss other observable manifestations of the FM transitions. 

The first order transition between the isotropic PM state and an anisotropic FM state is associated with the jump in the electronic density of states at the Fermi surface. It may be detected thus through a measurement of the electronic specific heat. The onset of the anisotropy in the orbital space may lead to anisotropic behavior of transport coefficients. This issue requires a separate study, since its involves specific scattering mechanisms. The Lifshitz transition between the ellipsoidal and toroidal Fermi surfaces does not lead to a discontinuity in the density of states. It leads, however, to its nonanalitic behavior,  $\nu(\tilde\alpha)=\nu - \nu_1 \sqrt{\tilde\alpha_3-\tilde\alpha}\, \theta(\tilde\alpha_3-\tilde\alpha)$, on the ellipsoidal side of the transition. This may also lead to  detectable features in the electronic specific heat.   

Besides thermodynamic or transport measurements, one may detect Friedel oscillations of the electronic density, created by an impurity.  Specifically, the emergence of the two Fermi momenta, $k_{F_1}$ and $k_{F_2}$, in the toroidal phase is reflected in a non-analytic behavior of a static charge susceptibility $ \chi(q) $ at points $q = 2k_{F_1}$, $q = 2k_{F_2}$, and $q = k_{F_1} \pm k_{F_2}$. Consequently, the spatial profile of the Friedel oscillations contains the corresponding harmonics, which can be observed using STM\@. 
The reconstruction of the Fermi surface may be also visible in the periods of the Shubnikov-de Haas magnetoresistance oscillations~\cite{Zhou2021Oct}. Indeed, the oscillation periods are associated with the areas of extremal cross sections of the Fermi surface, which change abruptly at all FM transitions, considered above.  Finally, the more delicate probes such as spin-resolved angle-resolved photoemission spectroscopy (ARPES), polarized neutron scattering, or the state-of-the-art nano-SQUID magnetometers~\cite{Zeldov2022} can reveal information about the shape of the Fermi surface and a local magnetic moment.  

\section{Concluding remarks}

We discussed how Rashba spin-orbit coupling of conduction band electrons, a result of the band mixing with spin-orbit-split valence bands, manifests itself in electron-electron interactions effects. Projection of a multi-band model onto the conduction band degrees of freedom leads to a spin-dependent renormalization of the electron density, as seen in Eq.~\eqref{Eq:H-int-coordinate}. Besides more familiar manifestations  in the single-particle dispersion relation (for a broken inversion symmetry), RSOI modifies the electron-electron interaction vertex. This phenomenon, known as the pair SOI, emerges without the need for external electric fields or inversion symmetry breaking, relying instead on internal fluctuating electric fields of the Coulomb interactions between (renormalized) conduction band electrons. 

We explored two consequences of PSOI\@. First, echoing the Kohn-Luttinger effect, PSOI induces $p-$wave superconducting pairing, driven purely by electron-electron interactions. This leads to distinct superconducting states in sufficiently clean materials. In 3D systems, the SOI constant $\alpha$ determines the nature of the superconducting order parameter, resulting in either nodal superconductivity along a spontaneously chosen direction or a node-less isotropic state. Strikingly, these phases mirror those in superfluid \isotope{He}{3}, drawing an unexpected parallel between these systems.

In 2D systems, the $p-$wave state is fully gapped. It consists of two decoupled condensates of spin up and down electrons, polarized along the normal direction.  The two condensate exhibit opposite Chern numbers. Their vortices (if spatially separated from each other) harbor Majorana states~\cite{Lutchyn2018May}. Stability of the latter depends on whether mutual interactions between vortices of the two condensates are repulsive (leading to stability) or attractive (leading to mutual annihilation). 

The potential for carrying spin-polarized supercurrents in these superconductors is particularly intriguing. This could be achieved through spin injection from a ferromagnet or by imbalancing the condensates via external magnetic fields or proximity effects. The concept of a superconducting spin-valve, controllable by magnetic fields or ferromagnetic leads, emerges as an exciting application.

The second consequence of PSOI is its role in facilitating Bloch ferromagnetism, linked with the symmetry breaking in both spin and orbital spaces. This effect transforms the Fermi surface of the polarized itinerant electrons into ellipsoidal or toroidal shapes, leading to genus-changing topological Lifshitz transitions as a function of the SOI magnitude or interactions strength. The transition may have a number of observable manifestations in thermodynamic and transport properties, as well as in the spectra of Friedel and Shubnikov-de Haas oscillations. Finally, the anisotropic Bloch ferromagnet may coexist with the $p-$wave superconductor.  If this is the case, the Lifshitz ellipsoidal to toroidal transition manifests itself as the transition between nodal and fully gapped superconducting states. 

The predicted effects of PSOI are highly sensitive to the Rashba constant, $\alpha$, which naturally drives the search for materials that i) exhibit giant Rashba effect, and ii) allow for ample tunability of the Rashba splitting by the external electric field. Promising candidates include thin layers of $\mathrm{Bi_2Se_3}$ and $\mathrm{KTaO_3}$~\footnote{In typical experiments, the layered materials are proximated by metallic gates. One then should take into account the emergent image-potential-induced PSOI, which leads to effects similar to those considered in the present paper~\cite{PhysRevB.95.045138,
2018arXiv180410826G,2019arXiv190409510G}.}, graphene on TMDs, van der Waals materials with heavy adatoms, and engineered heterostructures at the $\mathrm{LaAlO_3/SrTiO_3}$ interface~\cite{10.1063/5.0212170,Liu2023Feb}.

Yet, a significant gap persists in reliably estimating the Rashba coupling constant, particularly concerning the response of layered materials to in-plane electric fields. The Rashba coupling in promising 3D systems remains largely unknown. Many existing estimates rely on ARPES measurements, which mainly probe the Rashba effect from built-in i.e.~\textit{normal} electric fields. Crucially, direct \textit{ab initio} calculations of Rashba SOI from in-plane fields in 2D systems are lacking. This presents a dual challenge: while it complicates precise predictions of the effect's magnitude, it also opens an exciting new research avenue of searching for materials with a giant, tunable Rashba effect.

Recently, experimental studies have revealed specific features in electron transport within quantum wires at the $\mathrm{LaAlO_3/SrTiO_3}$ interface (see Ref.~\cite{Levy_2020} and references therein). These findings have been interpreted by their authors as signatures of bound electron pairs, potentially consistent with the pairing mechanism driven by PSOI that we explore in this paper~\cite{PhysRevB.104.125103}. Superconducting states observed in spin-orbit proximitized rhombohedral trilayer graphene could suggest the involvement of spin-dependent electron-electron interactions in pairing~\cite{young_2024}. However, further experimental and theoretical analysis is needed to definitively establish any connection to PSOI\@.

\begin{acknowledgments}
We thank A.~Chubukov, R.~Fernandes, M.~Glazov, D.~Gutman, S.~Kivelson, V.~Kozii, A.~Levchenko, D.~Maslov, Yu.~Oreg, P.~Littlewood, J.~Ruhman, V.~Sablikov, J.~Schmalian, B.~Shklovskii, G.~Vignale and P.~Wiegmann for helpful discussions. The work of Y.G.\ was supported by the Simons Foundation Grant No.~1249376;  A.K.\ was supported by the NSF Grant No.~DMR-2338819. Y.G.\ acknowledges the hospitality of the Kavli Institute for Theoretical Physics (KITP), Santa Barbara, supported by the National Science Foundation under Grants No.~NSF PHY-1748958 and PHY-2309135.
\end{acknowledgments}

\appendix
\section{Microscopic model of the RSOI}
\label{Sec:Microscopics}

Here a microscopic derivation of the RSOI Hamiltonian is outlined, with a focus on a momentum dependence of the SOI coupling for a particularly simple case of a 2D electron system with the inversion symmetry.
The approach is based on the symmetric Bernevig-Hughes-Zhang (BHZ) model~\cite{Bernevig1757}. This model is widely recognized for its capacity to describe various 2D systems with strong RSOI, across both topological and trivial phases.

The BHZ model is expressed in a four-band basis, denoted as $\mathcal{B}={\{\ket{\eup},\ket{\hup},\ket{\edn},\ket{\hdn}\}}^{\intercal}$. Within this basis, the states $\ket{\eup}$ and $\ket{\edn}$ consist of the electron- and light-hole band states, each carrying an angular momentum projection of $m_J = \pm 1/2$. Conversely, the states $\ket{\hup}$ and $\ket{\hdn}$ correspond to the heavy-hole states, with an angular momentum projection of $m_J = \pm 3/2$. The Hamiltonian governing the envelope wave-function in this context is as follows
\begin{equation}
    \hat{H}_{\mathrm{BHZ}} = 
    \begin{bmatrix}
        m(\hat{p})      & A \hat{p}_+ & 0 &     0\\
        A \hat{p}_-     & -m(\hat{p}) & 0          &     0\\
        0 & 0     & m(\hat{p})       & -A \hat{p}_-\\
        0         &0      & -A \hat{p}_+     & -m(\hat{p})
    \end{bmatrix}\,.
\end{equation}
Here 
\begin{equation}
    \label{Eq:m-gap}
    m(\hat{p})= m + B \hat{p}^2\,,
\end{equation}
with $m$ the band gap, $B$ the bare dispersion curvature,  and $A$ the band hybridization parameter. Then, $\hat{p}$ is the momentum operator, $\hat{p}_{\pm} = \hat{p}_x \pm i \hat{p}_y$.

The BHZ model inherently incorporates a strong SOI, premised on the assumption of an infinitely large spin-splitting in the valence band.
Since the spin sectors are decoupled it is possible to treat them separately by considering a $2\times2$ Hamiltonian
\begin{equation}
    \hat{H}_0(\hat{p}) = m(\hat{p}) \tau_3 + A (\hat{\bm{p}}\cdot \hat{\bm{\tau}})\,,
\end{equation}
with $\hat{\tau}_i$ the pseudospin (subband) Pauli matrices. The Hamiltonian can be diagonalized
\begin{equation}
    \hat{\mathcal{H}}_0(\hat{p}) = \hat{U}(\hat{p})\hat{H}_0 (\hat{p}) \hat{U}^+(\hat{p}) 
\end{equation}
by a unitary Foldy-Wouthuysen transformation
\begin{equation}
            \label{Eq:U}
    \hat{U}(\hat{p}) = \cos \theta(\hat{p}) + \sin \theta(\hat{p}) \frac{(\hat{\bm{p}}\cdot \hat{\bm{\tau}})}{p}\tau_3
\end{equation}
to acquire the form
\begin{equation}
    \hat{\mathcal{H}}_0(\hat{p}) = \varepsilon(\hat{p}) \tau_3\,,
\end{equation}
with the band dispersion
\begin{equation}
\label{Eq:BHZ_dispersion}
    \varepsilon(\hat{p}) = \sqrt{m^2(\hat{p}) + A^2 \hat{p}^2}\,,
\end{equation}
and
\begin{equation}
    \tan 2\theta(\hat{p}) = -A \frac{\hat{p}}{m(\hat{p})}\,.
\end{equation}

Applying the unitary transformation to the Hamiltonian
\begin{equation}
    \hat{H}(\hat{p}) = \hat{H}_0(\hat{p}) - e \varphi (\hat{\bm{r}}), 
\end{equation}
that describes the interaction with a scalar potential, one should take into account that the momentum-dependent operators $\hat{U}(\hat{p})$ do not commute with a position-dependent $\varphi (\hat{\bm{r}})$. The commutator terms generate the operator
\begin{equation}
    \hat{\mathcal{V}}(\hat{p}) = \frac{\sin^2 \theta(p)}{p^2} \hat{\bm{\tau}} \cdot (-e \grad_{\bm{r}} \varphi \times \hat{\bm{p}})\,.
\end{equation}

Within the low-energy approximation one focuses on the electron dynamics in the conduction band by declaring the spinor components in the valence band to be small and keeping only the $\hat{\mathcal{V}}_{11}$ matrix element pertaining to the conduction band. Combining the terms  $\hat{\mathcal{V}}_{11}^{(s)}$ from both spin sectors  the following generalized RSOI Hamiltonian in the reduced conduction band basis is obtained,
\begin{equation}
    \label{Eq:bhzrsoi}
    \hat{H}_{\mathrm{RSOI}}(\hat{p}) = \frac{\sin^2 \theta(p)}{p^2} \hat{\bm{\sigma}} \cdot (-e \grad_{\bm{r}} \varphi \times \hat{\bm{p}})\,,
\end{equation}
with $\hat{\bm{\sigma}}$ the real spin Pauli vector. This gives rise to the momentum-dependent SO coupling of the form
\begin{equation}
    \label{Eq:AlphaBHZ}
    \alpha_{pp} = \frac{e}{2 p^2}\left(1-\frac{m(p)}{\varepsilon(p)}\right)\,.
\end{equation}
In the long-wave limit it is consistent with an earlier result of Ref.~\cite{Rothe_2010},
\begin{equation}
\label{Eq:AlphaBHZ_1}
    \alpha_{pp} = \frac{e A^2}{4 m^2(p)}\,,
\end{equation}
while at large momenta, when $\varepsilon(p) \gg m(p)$, the SO coupling asymptotically behaves as
\begin{equation}
    \label{Eq:saturation}
    \alpha_{pp} \sim \frac{e}{2 p^2}\,.
\end{equation}

\section{Self-consistent HF solution}
\label{Sec:Torus}
Here, the toroidal Fermi surface is demonstrated to be a self-consistent solution of Eq.~\eqref{Eq:MF-polarized}. The derivation proceeds in two steps. First, the effective dispersion in Eq.~\eqref{Eq:MF-polarized} is found assuming that the electrons occupy the toric region in the momentum space limited by the Fermi surface of Eq.~\eqref{Eq:Oblate_FS}. Second, the isoenergetic surface of the resulting HF potential is verified to be that of Eq.~\eqref{Eq:Oblate_FS} under the proper choice of parameters, which results in self-consistency conditions.

A change of variables $q_{\parallel}^2 = p'^2_{\parallel} +\delta^2$, $q_z = p'_z$ with $\delta = (\Lambda - \zeta)/4 K$ transforms Eq.~\eqref{Eq:Oblate_FS} to the canonical equation of a torus,
\begin{equation}
    \label{torus}
    {(q^2 + K^2 -k^2)}^2 = 4 K^2 q_{\parallel}^2\,,
\end{equation}
where
\begin{equation}
    \label{Eq:def-k}
        k^2 = {(K - \delta)}^2 - \zeta\,.
\end{equation}
The torus interior, $\mathcal{D}$, can be parametrized by
\begin{equation}
    \begin{split}
        q_x &= (K + \eta \cos \theta)\cos \phi\,,\\
        q_y &= (K + \eta \cos \theta)\sin \phi\,,\\
        q_z &= \eta \sin \theta\,,\\
    \end{split}
\end{equation}
with $\eta \in (0,k)$, $\phi \in (0,2\pi)$, and $\theta \in (0,2\pi)$. The Jacobian of the transformation is $J(\eta,\theta) = \eta (K + \eta \cos \theta)$.

The momentum-dependent factor that enters Eq.~\eqref{Eq:MF-polarized} is
\begin{equation}
    \mathcal{M}_z^2(\bm{p},\bm{p}') = {[\bm{p} \times \bm{p'}]}^2_z = p^2_{\parallel} p'^2_{\parallel} \sin^2 \widehat{\bm{p}_{\parallel} \bm{p}'_{\parallel}}\,.
\end{equation}
Due to the axial symmetry of the problem, the $\sin^2 \widehat{\bm{p}_{\parallel} \bm{p}'_{\parallel}}$ averages to $1/2$ when integrated over the azimuthal angle.
The PSOI contribution to the HF potential of Eq.~\eqref{Eq:MF-polarized} takes the form
\begin{equation}
    \label{Eq:MF-ferro-integrals}
    \begin{split}
        V_{\mathrm{PSOI}}(\bm{p}) =&{} - \frac{\alpha^2\mathcal{U}_0 \Lambda^2}{\Lambda + p^2} \int \frac{d^3 p'}{{(2 \pi)}^3} \frac{{[\bm{p} \times \bm{p'}]}^2_z}{\Lambda + p'^2}\\
        =&{} - \frac{\alpha^2\mathcal{U}_0 \Lambda^2 p_{\parallel}^2}{\Lambda + p^2} \frac{\Xi}{2}\,,
    \end{split}
\end{equation}
where
\begin{widetext}
\begin{equation}
    \begin{split}
        \Xi =&{} \int \frac{d^3 p'}{{(2 \pi)}^3} \frac{p'^2_{\parallel}}{\Lambda + p'^2}
        = \int_{\mathcal{D}} \frac{d\eta d\theta d\phi}{{(2 \pi)}^3} J(\eta,\theta)\frac{{(K + \eta \cos \theta)}^2 - \delta^2}{\eta^2 +2 K \eta \cos \theta + K^2 + \Lambda - \delta^2}\\
        ={}& \frac{1}{96 \pi K^3}\Big[ k^6-k^4 \left(3 \delta ^2+\varkappa +3 K^2-3 \Lambda \right)+k^2 \left(\left(\delta ^2-\Lambda \right) \left(3 \delta ^2+2 \varkappa -3 \Lambda \right)+21 K^4+2 K^2 (\varkappa -6 \Lambda )\right)\\
        &{}+\left({\left(\delta ^2-\Lambda \right)}^2+K^4-2 K^2 \left(\delta ^2+5 \Lambda \right)\right) \left(-\delta ^2-\varkappa +K^2+\Lambda \right) \Big]\,,
    \end{split}
\end{equation}
\end{widetext}
with
\begin{equation}
    \varkappa \equiv \sqrt{\left[{(K+k)}^2+\Lambda -\delta^2\right] \left[{(K-k)}^2+\Lambda -\delta^2 \right]}\,.
\end{equation}
Consequently, the total HF potential of Eq.~\eqref{Eq:MF-polarized} acquires the form
\begin{equation}
    \label{Eq:M-Field-sc}
    V_{\bm{p}}^{\mathrm{HF}} = \frac{ p^2}{2m} - e^2 \mathcal{U}_0 n- \frac{\alpha^2\mathcal{U}_0 \Lambda^2 p_{\parallel}^2}{\Lambda + p^2} \frac{\Xi}{2}\,.
\end{equation}

The Fermi surface of $V_{\bm{p}}^{\mathrm{HF}} = E_F$ for the electron dispersion of Eq.~\eqref{Eq:M-Field-sc} has indeed the assumed form of Eq.~\eqref{Eq:Oblate_FS}, provided that the following self-consistency conditions hold:
    \begin{align}
        K^2 &= \frac{m \alpha^2\mathcal{U}_0 \Lambda^2}{4 } \Xi\,,\\
        \zeta &= -2m (E_F + e^2 \mathcal{U}_0 n)\,,
    \end{align}
which should be supplemented by Eq.~\eqref{Eq:def-k} and the relation $k^2 K = 4 \pi n$ between the Fermi sea volume and electron concentration. The system of these four equations allows one to find the three quantities: $K$, $k$, and $\zeta$, which determine the geometry of the Fermi surface, as well as the Fermi energy $E_F$. 

\section{2D Rashba Ferromagnets}
\label{Sec:Ferro-2D}
Here the effects of PSOI are explored in 2D electron systems. To begin, consider a particular case of a system symmetric with respect to the inversion of the normal $z \to -z$. Within this framework, the PSOI emerges solely due to the in-plane Coulomb fields, created by interacting electrons. The symmetric 2D structures represent a fascinating object of research in their own right, as they are  realized through freely suspended 2D layers, with a focus on amplifying the effects of e-e interactions~\cite{Pogosov_2022}.

The expressions for the effective electron dispersion and exchange energy presented in Section~\ref{Sec:TPT} also apply here, with the integration running over the 2D momentum space. The analysis is performed within the BHZ model~\cite{Bernevig1757}, with the spin-orbit coupling given by Eq.~\eqref{Eq:AlphaBHZ_1}, and kinetic energy of Eq.~\eqref{Eq:BHZ_dispersion}.

The interaction in 2D layers of Rashba materials is governed by the Rytova-Keldysh potential~\cite{rytova,keldysh1979coulomb}
\begin{equation}
    \mathcal{U}_{\bm{q}} = \frac{2 \pi}{q (1+ q l)}\,,
\end{equation}
where the length $l$ is a phenomenological parameter of the theory, which generally speaking should be determined from the first-principles calculations~\cite{RevModPhys.90.021001}. It can be estimated from the layer polarizability as $l = \epsilon d/2$, where $d$ stands for the layer thickness, and $\epsilon$ is the in-plane component of the dielectric tensor of the corresponding bulk material.

Spin polarization  is controlled by the balance between the kinetic and exchange interaction (both Coulomb and PSOI) energies. The degree of the spin polarization is found by identifying the minimum of the total HF energy. As the three contributions display distinct density dependencies~\cite{review_jetp}, it is anticipated that the lowest energy state will be FM in two regions: at a low density as a result of Coulomb exchange and at a high density due to PSOI\@.

\begin{figure}[htb]
	\includegraphics[width=0.9\linewidth]{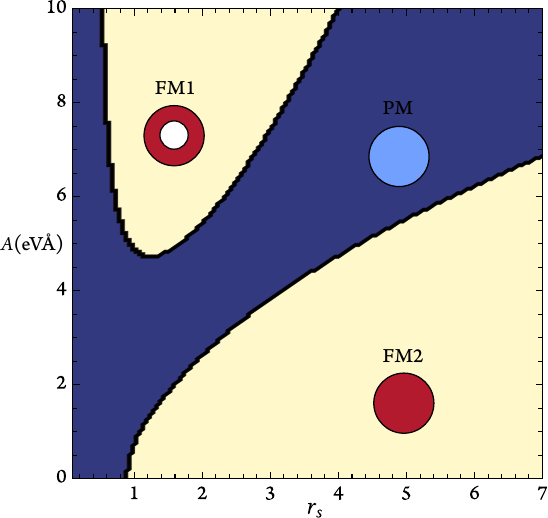}
	\caption{The phase diagram for the spin polarization of a 2D electron gas with PSOI\@. The system parameters are those typically found in 2D Rashba materials~\cite{manchon2015new}: $m = \SI{0.01}{eV}$, $B = \SI{50}{{\angstrom^2} eV}$, $d = \SI{50}{\angstrom}$, $\epsilon = 20$. \label{Fig:diag2D}}
\end{figure}

These expectations are confirmed by Fig.~\ref{Fig:diag2D}, which presents the phase diagram of system magnetization vs.  BHZ velocity $ A $, which serves as the  magnitude of the PSOI, and $ r_s $ values~\footnote{The parameter $r_s$, defined through the Bohr radius, typically describes a Galilean-invariant electron system that interacts via a power-law Coulomb potential. It is clear that neither of these characteristics apply to our system. However, we still need to establish some scale to investigate the dependence of the HF energy on the electron concentration. Thus we  formally define the Bohr radius $a_B$ for the Coulomb interaction within the bulk material, characterized by the dielectric constant $\epsilon$. This definition is made under the assumption of an effective electron mass, as determined by the bare band curvature $B$.}.

One identifies two distinct, fully spin-polarized phases. A new region of the FM phase emerges at lower $r_s$ values when the PSOI magnitude surpasses a certain critical threshold. Its Fermi surface exhibits a shape of annulus $k_{F_1} \le k \le k_{F_2}$, which can be viewed as a 2D projection of the toroidal Fermi-surface discussed in Section~\ref{Sec:Ferro}.

This result  is seemingly at odds with the conventional perspective on Bloch ferromagnetism, which is expected to only occur at sufficiently low electron densities. Such transition, driven by the Coulomb exchange, is indeed present as FM2 phase in Fig.~\ref{Fig:diag2D} in a FM region of high $ r_s $. It exhibits a circular Fermi surface.  Conversely, the annular FM1 phase emerges in the regime of a dense electron gas,  where the PSOI interactions are strong due to their strong momentum dependence.
Both FM phases exhibit spin polarization normal to the plane, with the easy axis set by the PSOI\@. This is also different  from the conventional 2D Bloch ferromagnetism, where the spin rotation symmetry can spontaneously break in any direction.

The emergence of the annular FM phase is due to the moat-band dispersion created by PSOI\@. Notice that the single-particle Rashba effect, due to a normal electric field $\mathcal{E}_z$, also leads to a similar moat-band. In the latter scenario, the electron system displays a momentum-locked chiral spin texture. In the annular FM phase, on the other hand, electrons exhibit complete spin polarization in the $ z $-direction.

In both FM1 and FM2 phases, electrons remain unaffected by the single-particle RSOI, $H_{\mathrm{RSOI}} = \alpha (\bm{\sigma} \times \bm{\mathcal{E}}_z)\cdot \bm{p}$, since the magnetization, $\bm{\sigma}$, is colinear with the external field. Only when the electric field surpasses a certain critical magnitude of the order of $\mathcal{E}_c \simeq v_F/\alpha$, the $z$-polarized FM phases lose their stability.  A transition to an alternative spin-polarized phase, potentially exhibiting nematic order, is plausible~\cite{PhysRevB.85.035116,PhysRevB.89.155103,PhysRevB.90.235119} at a larger external field. 

\bibliography{paper}

\end{document}